\documentclass[12pt,preprint]{aastex}

\newcommand{\kms}{km~s$^{-1}$}
\newcommand{\Htwo}{\mbox{H$_{2}$}}

\shorttitle{The HD\,192639 line of sight}
\shortauthors{Sonnentrucker et al.}

\begin{document}


\title{Abundances and Physical Conditions in the Interstellar Gas toward HD\,192639.\altaffilmark{1}}


\author{P. Sonnentrucker\altaffilmark{2}, S. D. Friedman\altaffilmark{2}, D. E. Welty\altaffilmark{3}, D. G. York\altaffilmark{3} and T. P. Snow\altaffilmark{4} }

\altaffiltext{1}{``Based on observations made with the NASA/ESA Hubble Space Telescope, obtained from the Data Archive at the Space Telescope Science Institute, which is operated by the Association of Universities for Research in Astronomy, Inc., under NASA contract NAS 5-26555. These observations are associated with proposal \#8241.''}
\altaffiltext{2}{Department of Physics and Astronomy, Johns Hopkins University, 3400 North Charles Street, Baltimore, MD 21218}
\altaffiltext{3}{Department of Astronomy and Astrophysics, University of Chicago, 5640 South Ellis Avenue, Chicago, IL 60637.}
\altaffiltext{4}{Center for Astrophysics and Space Astronomy, University of Colorado, Campus Box 389, Boulder, CO 80309}



\begin{abstract}
We present a study of the abundances and physical conditions in the interstellar gas toward the heavily reddened star HD\,192639 [$E_{(B-V)}$ = 0.64], based on analysis of {\it FUSE} and {\it HST}/STIS spectra covering the range 
from 912 to 1361 \AA. This work constitutes a survey of the analyses that can be performed to study the interstellar gas when combining data from different instruments. Low-velocity ($-$18 to $-$8 km s$^{-1}$) components are 
seen primarily for various neutral and singly ionized species such as 
\ion{C}{1}, \ion{O}{1}, \ion{S}{1}, \ion{Mg}{2}, \ion{Cl}{1}, \ion{Cl}{2}, 
\ion{Mn}{2}, \ion{Fe}{2} and \ion{Cu}{2}.
Numerous lines of H$_2$ are present in the {\it FUSE} spectra, with a 
kinetic temperature for the lowest rotational levels $T_{01}$ = (90 $\pm$ 10) K.
Analysis of the \ion{C}{1} fine-structure excitation implies an average local 
density of hydrogen $n_{\rm H}$ = (16 $\pm$ 3) cm$^{-3}$.
The average electron density, derived from five neutral/first ion pairs under 
the assumption of photoionization equilibrium, is $n_e$ = (0.11 $\pm$ 0.02) 
cm$^{-3}$. The relatively complex component structure seen in high-resolution spectra of \ion{K}{1} and \ion{Na}{1}, the relatively low average density, and the measured depletions all suggest that the line of sight contains a number of diffuse clouds, rather than a single dense, translucent cloud. Comparisons of the fractions of Cl in \ion{Cl}{1} and of hydrogen in molecular form suggest a higher molecular fraction, in the region(s) where \Htwo\, is present, than that derived considering the average line of sight. In general, such comparisons may allow the identification and characterization of translucent portions of such complex lines of sight. The combined data also show high-velocity components near $-$80 km s$^{-1}$ for various species which appear to be predominantly ionized, and may be due to a radiative shock. A brief overview of the conditions in this gas will be given.
\end{abstract}


\keywords{ISM: abundances, atoms---ISM: clouds, molecules---stars: individual (HD\,192639)}


\section{Introduction}

The {\it Far Ultraviolet Spectroscopic Explorer} ({\it FUSE}) allows us to study \Htwo\, directly in the ground state absorption lines. Its wavelength coverage (912-1187 \AA) includes numerous electronic transitions from the strong \Htwo\, Werner and Lyman bands as well as from a wealth of heavy elements crucial for diagnosing the temperature, abundance, depletion and density of the interstellar gas. Furthermore, the {\it FUSE} sensitivity permits the observation of fainter stars with higher extinctions than previously possible with far-UV instruments \citep{moos00}. Hence, {\it FUSE} offers a unique opportunity to study the properties of clouds with visual extinction $A_{V}$ greater than 1 magnitude (mag), defined as translucent clouds. Little is known about the UV absorption-line properties of such clouds since their observation was technically difficult until the {\it FUSE} era. What are the dominant constituents of translucent clouds apart from molecular hydrogen? How severe is the depletion in such clouds? At what temperature and density is the gas in such clouds? What characterizes the transition from diffuse to translucent clouds? \\

The {\it FUSE} translucent cloud program was designed to answer these questions. It involves more than 30 stars chosen
for their values of $A_{V}$ ranging from 1 to 5 mag in an effort to distinguish translucent clouds from ensembles of diffuse clouds, to determine abundances, depletions, and physical conditions in those clouds, and to search for clouds presenting properties intermediate between diffuse and translucent (see HD\,110432 in Rachford et al. 2001a). HD\,192639 is among the stars selected in the {\it FUSE} translucent cloud program for its extinction ($A_{V} \sim $2), typical of that class of clouds.
This star was also observed with {\it HST} in ``The Snapshot Survey of Interstellar Absorption Lines'' conducted by J. Lauroesch (P8241). \\

HD\,192639 is an O7.5 Ib(f) \citep{walborn} star bearing a reddening E$_{(B-V)} =$ 0.64. This star is a member of the Cyg OB1 Association, one of the most luminous OB 
associations in the solar neighborhood.  The association is centered at
$(l^{II},b^{II}) = (76^{\circ},1^{\circ})$ at an estimated distance of 1.8 kpc 
\citep{Stl91}.
Previous studies of the Cygnus region have shown it to be dynamically active, with shocked gas and related \ion{H}{2} regions being detected at high and intermediate velocity toward Cyg OB1 and OB3 \citep{phi84,Stl91,loz98a}. Equivalent width estimates were obtained from {\it IUE} data for a few heavy elements \citep{phi84, Stl91}, and molecular hydrogen (\Htwo\,) column densities were indirectly derived from infrared measurements which are sensitive only to the highest density regions.\\

We report here the far ultraviolet (FUV) study of the potentially translucent cloud line of sight toward HD\,192639. We present a survey of the analyses that can be performed when combining data from different instruments. Such combinations allow us to detect a wide range of species in various ionization states and to perform detailed studies of the local conditions in the ISM gas. This study focuses on the detailed analysis of the cold interstellar gas toward HD\,192639 but will also briefly give an overview of the other gas components detected in the combined datasets along this line of sight. This work comprises the analysis of the FUV absorption lines included in the combined {\it FUSE} and archival {\it HST}/STIS datasets covering the wavelength range 912-1361 \AA. The resolutions of 18 \kms\, ({\it FUSE} low-resolution aperture)
and 2.75 \kms\, (STIS E140H), supported by high-resolution optical \ion{K}{1} and \ion{Na}{1} data ($\sim$1.5-1.6 \kms), allowed us to derive accurate column densities for many important gas-phase species, among which are \ion{C}{1}, \ion{C}{1*}, \ion{C}{1**}, \ion{N}{2}, \ion{O}{1}, \ion{O}{1*}, \ion{O}{1**},  \ion{Mg}{2}, \ion{Si}{2}, \ion{S}{1}, \ion{Cl}{2}, \ion{Mn}{2}, \ion{Fe}{2} and \Htwo. The data reduction is described in \S\, 2. A study of the velocity structure toward HD\,192639 is found in \S\, 3. The physical properties of the low-velocity gas in terms of pressure, temperature, density and depletion are derived and discussed in \S\, 4. A brief study of the high-velocity gas detected with the combined data toward HD\,192639 in the \ion{C}{2}, \ion{C}{2*}, \ion{N}{2}, \ion{N}{2*}, \ion{Si}{2}, \ion{Si}{2*} and \ion{S}{2} lines is presented in \S\, 5 and the main results of this work are summarized in \S\, 6.

\section{Data processing and analysis}

\begin{figure}
\caption{The upper left panel contains a sample of \Htwo\, lines from the LVCs detected in the {\it FUSE} LiF1A spectrum. The low $J$ lines are labeled, correspondingly in wavelength space. The higher $J$ levels seen in this panel are labeled from 1 to 5 and represented in each of the 5 remaining panels using heliocentric velocities. The tick marks indicate the position of the 5 components forming the LVCs. \label{hdeux}}
\plotone{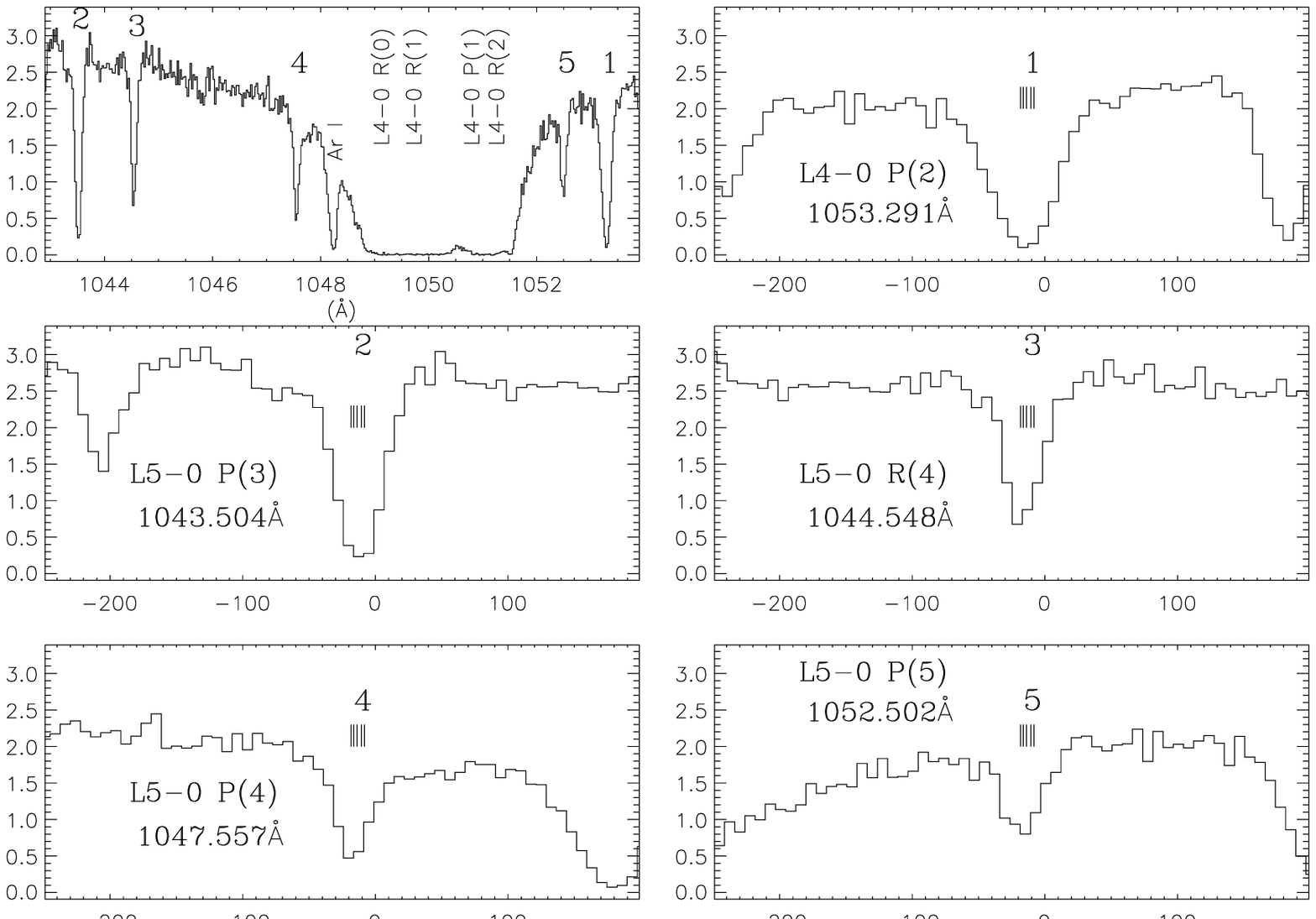}
\end{figure}

{\it FUSE} observed HD\,192639 (Program P1162401) in June 2000, in time tag 
mode \citep{sahnow} for a total of 4834s with the low resolution aperture (LWRS), covering
 the wavelength range 912-1187 \AA. The data resolution is about 0.062 \AA\ ({\rm FWHM}, corresponding to $\sim$18 \kms\, and $\sim$9 pixels). 
No detector burst activity was found in either of the two observations \citep{sahnow}.
The data were processed with version 1.8.7 of the CALFUSE pipeline which 
included geometric distortion correction, Doppler correction to set the heliocentric rest-frame, dead-time correction and wavelength calibration \citep{kruk, sahnow}. The exposures were co-added after cross-correlating and shifting the spectra with respect to the brighter exposure, for each channel. The shifts are 8 pixels maximum. The LiF1B spectrum was excluded from the data analysis due to the presence of a well known detector artifact causing an artificial flux deficiency in this segment \citep{kruk}. Figure \ref{hdeux} presents a sample of the molecular hydrogen lines used to derive the column densities in the rotational states $J=$ 0 to $J=$ 5. One of the windows used to fit each transition (see \S\, 4.2) is shown in the upper left panel of Fig. \ref{hdeux}. The remaining 5 panels show selected lines for each rotational transition in the heliocentric rest-frame. The spectra used for the data analysis were binned by 4 pixels (less than one resolution element) in order to increase the signal-to-noise (S/N) ratio without degrading the resolution. The co-added spectra have S/N of about 10 per resolution element in the LiF1A spectrum and up to 15 per resolution element in the LiF2A spectrum. Weak lines such as those from \ion{Cl}{2}, \ion{C}{1**} or \ion{Fe}{2} were assumed to be positively detected (and not the result of fixed pattern noise contamination) when the lines were present in more than one channel.\\

The STIS data for HD\,192639 were obtained for ``The Snapshot Survey of Interstellar Absorption Lines'' (P8241) conducted by J. Lauroesch using the FUV MAMA detector and the 0.''2 $\times$ 0.''2 aperture. The E140H echelle grating provides a resolution of 2.75 \kms\, and spans the wavelength range between 1150-1361 \AA, therefore complementing the {\it FUSE} 
bandpass. The 49 extracted orders were corrected for scattered light contamination following the procedure described in \citet{Howk00}.\\  

\begin{figure}
\caption{Absorption from the trace neutral species in the LVCs. The wavelengths correspond to the element indicated by the tick marks. The source of the data is indicated at the bottom of each panel: (I) for {\it IUE} data, (F) for {\it FUSE} data, (O) for optical data and (S) for STIS data. The tick marks indicate the position of the 5 components forming the LVCs. \label{vel}}
\plotone{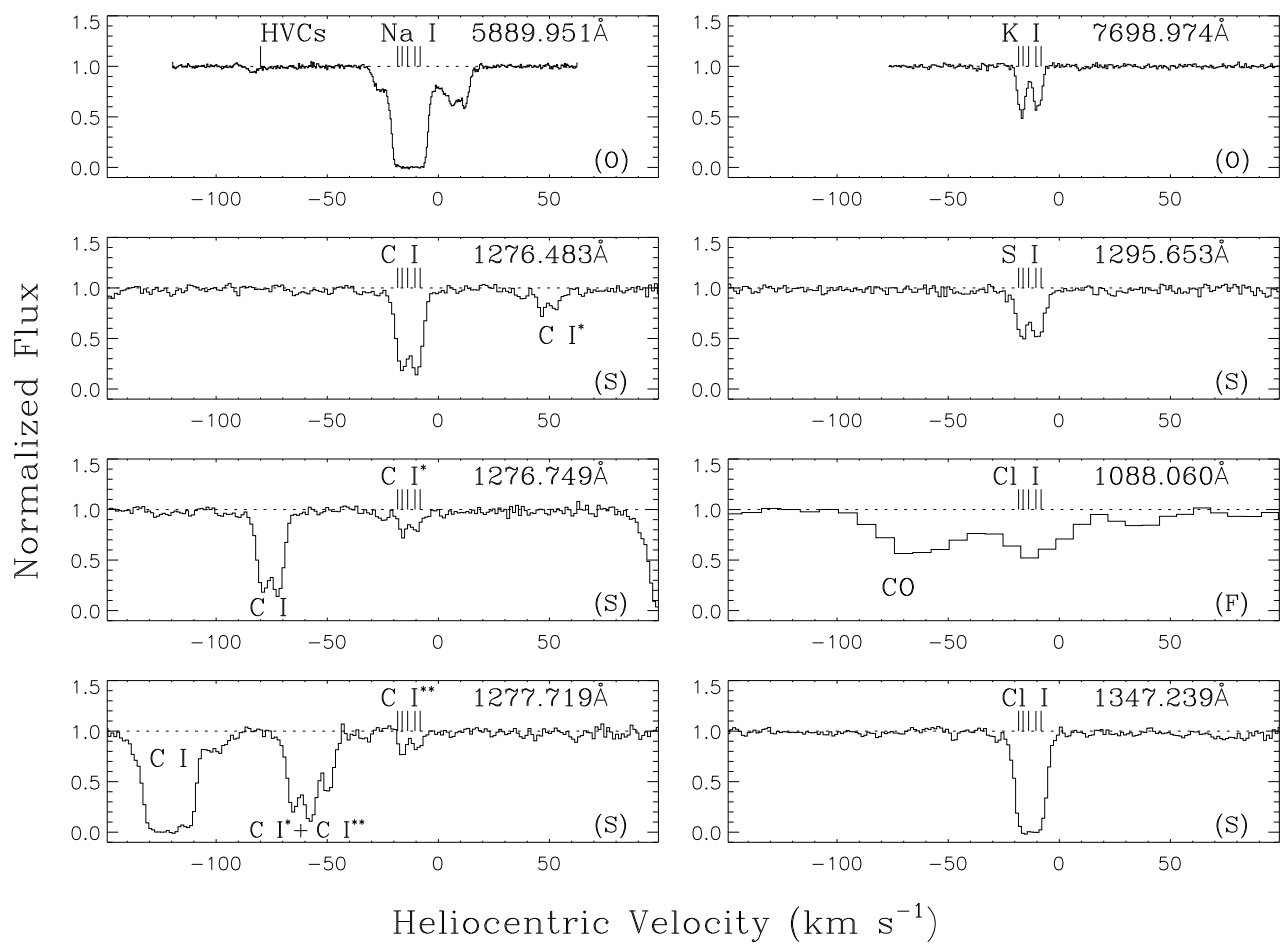}
\end{figure}

High-resolution (FWHM $\sim$ 1.5--1.6 km s$^{-1}$) spectra of \ion{K}{1}, 
\ion{Na}{1}, \ion{Ca}{1}, CH, and CH$^+$ were obtained with the Kitt Peak 
coud\'{e} feed telescope in 1998.
A detailed discussion of the reduction and analyses of these spectra --- and of similar spectra for other stars in the {\it FUSE} translucent cloud survey 
(Rachford et al. 2001b) --- will be given by Welty et al. (2002b in prep.).
The spectra of \ion{K}{1} and \ion{Na}{1} are shown in Fig. \ref{vel} along with lines detected with STIS and {\it FUSE}.\\

HD\,192639 was observed with the echelle spectrographs on board the {\it IUE} satellite at a resolution of 0.1-0.2 \AA\ for a total of 226 min. Figure \ref{veld} shows the \ion{C}{4} and \ion{Si}{4} lines detected in the short wavelength image (SWP 9493) which resulted from an integration of 136 min \citep{phi84}. The data shown here were retrieved from the MAST archival database and used without any additional processing. \\

The absorption lines used in the following analysis are shown in Figs. \ref{hdeux} through 6. Each species of interest is indicated by a series of 5 tick marks positioned at the velocities of each one of the 5 components detected in the low-velocity gas (LVCs, see \S\, 3). The corresponding wavelength of each species is noted at the top right of each panel. No shifts were necessary to align the STIS data with the ground based data. The {\it FUSE} data were shifted with respect to the STIS data by $-$16 \kms\, (LiF1A), $-$10 \kms\, (LiF2A) and +31 \kms\, (SiC2B; SiC1A). A shift of $-$10 \kms\, was applied to the {\it IUE} data to align the \ion{S}{2} and \ion{Si}{2} lines with the corresponding STIS lines. If high-velocity components (HVCs) are detected (see \S\, 3), an additional tick mark is drawn at their average position ($-$80 \kms\,). All velocities are heliocentric. When other absorption lines are present in the velocity range considered, those elements are noted below the spectrum. The source of the data is indicated at the bottom right of each panel as follows: (I) for {\it IUE} data, (F) for {\it FUSE} data, (O) for optical data and (S) for STIS data. The continua for the lines were normalized to unity, in Figs. 2 to 6, by applying low-order polynomials to the spectral regions adjacent to each absorption line.\\




\section{Line-of-sight velocity structure}

The detailed interstellar component velocity structure was inferred from the 
high-resolution \ion{K}{1} and \ion{Na}{1} spectra.
Fits to the \ion{K}{1} line profile indicate (at least) five components, at 
velocities from $-$18 to $-$8 km s$^{-1}$, which comprise 
the three-component groups discernible in the lower resolution STIS spectra of 
species such as \ion{C}{1}, \ion{O}{1}, \ion{S}{1}, \ion{Mg}{2}, or \ion{Ni}{2} (Figs. \ref{vel}, \ref{velb} and \ref{velc}). For T$\sim$100 K (see below), the \ion{K}{1} $b$-values (0.8-1.2 \kms) suggest that the line broadening is dominated by
turbulence for each of the components.
These three-component groups seen in STIS contain predominantly neutral (\ion{H}{1}) gas, most of the \ion{H}{1} and H$_2$ along this line of sight, and 
will be referred to as the Low-Velocity Components (LVCs).

Some of the stronger UV lines (e.g., of \ion{C}{2}, \ion{Si}{2}, and 
\ion{Si}{3}) reveal absorption from high-velocity gas near $-$80 km s$^{-1}$; 
at least two components can be discerned in the profiles of \ion{Si}{2}* 
and \ion{S}{2}.
Spectra from {\it IUE} show absorption from yet higher ions (\ion{C}{4}, 
\ion{Si}{4}, \ion{Al}{3}, \ion{S}{3}) at those velocities, indicating that 
the High-Velocity Components (HVCs) are predominantly ionized 
(Phillips et al. 1984).
We note, however, that high-velocity absorption is also seen for \ion{N}{1}, and \ion{Na}{1} --- so that some neutral gas is present as well.

While we focus in the next sections on the abundances and physical conditions in the LVCs and briefly discuss the HVCs, we note that several relatively weak components, at velocities from $-$27 km s$^{-1}$ 
to +12 km s$^{-1}$, are seen in \ion{Na}{1} (and in the UV lines of various 
other species). Additional intermediate-velocity components, extending to 
about $-$60 km s$^{-1}$, are present in the stronger UV lines of \ion{C}{2}, 
\ion{N}{1}, \ion{O}{1}, and \ion{Si}{2}.
These components likely contain mixtures of neutral and ionized gas; the 
absorption at +10 km s$^{-1}$ may largely arise in an \ion{H}{2} region 
(Phillips et al. 1984; St-Louis \& Smith 1991).

\section{The Low-Velocity Components (LVCs)}

The equivalent widths of the heavy elements measured over the velocity range $-$30 to 0 \kms\, are summarized in Table \ref{ewlow}. Wavelengths and $f$-values were taken from (Morton 2001 in prep.) unless noted otherwise in Table 1.
The column densities of the species present in both {\it FUSE} and STIS datasets were derived using up to four different methods depending on the type and number of transitions available; 1) a curve of growth (COG) analysis, 2) an apparent optical depth (AOD) analysis for optically thin or slightly saturated lines, 3) Equivalent Width (EW) analysis for optically thin lines and 4) a profile-fitting (FIT) analysis based on the ``Owens'' code described by \citet{lemoine}. When multiple methods were used and showed consistent results, we calculated the weighted mean of the column densities, weighting each column density by its error for a given species. The results and associated 1$\sigma$ errors are reported in Table \ref{low}. The methods used to derived the column densities are indicated in column (3) of Table \ref{low}.

\subsection{Component analysis of the LV gas}

Lines from a number of neutral and ionized species present in the STIS spectra, \ion{C}{1} and \ion{S}{1}, exhibit profiles similar to that of \ion{K}{1}. Because the profile analysis of the high-resolution \ion{K}{1} lines shows that the gas motion in the LVCs is dominated by turbulence at the typical temperature of the cold interstellar medium (100 K), we used the \ion{K}{1} component structure to fit the STIS \ion{C}{1}, \ion{C}{1*}, \ion{C}{1**}, \ion{O}{1}, \ion{Mg}{2}, \ion{S}{1}, \ion{Cl}{1}, \ion{Mn}{2}, \ion{Fe}{2}, \ion{Ni}{2}, \ion{Cu}{2} and \ion{Ge}{2} profiles. The number of components and $b$-values were held fixed, while the component column densities and the continuum were allowed to vary. Because the five components detected in
the high-resolution \ion{K}{1} spectra are both narrow ($b$ $\sim$ 
0.8--1.2 km s$^{-1}$) and separated from each other by only 2-3
km s$^{-1}$, they are only partially resolved in the STIS spectra.
In such cases, even the relatively small uncertainties in velocity zero 
point in the STIS spectra can lead to large uncertainties in the column 
densities of some individual components.  While the total line-of-sight 
LV column densities are fairly well determined, some of the individual  
LV component column densities listed in Table 2 thus are known only to 
within a factor of 2.  Given those uncertainties, it is difficult
to distinguish differences in the abundances, depletions, pressures,
and densities among the individual LV components.
The gas in all five components seems to exhibit similar conditions suggesting that each component contributes similarly to the total pressure, density, abundances and depletions over the entire LVCs. In light of the generally good correlation between $N$(CH) and $N$(H$_2$) in interstellar clouds (Danks, Federman, \& Lambert 1984), the roughly constant ratio $N$(\ion{K}{1})/$N$(CH) found for the five LV components suggests that each component also contributes similarly to H$_2$ ($J=$ 0 to 2) and (perhaps) to CO.\\

The low resolution (FWHM $\sim$ 18 km s$^{-1}$) and small-scale wavelength
uncertainties in the {\it FUSE} spectra preclude detailed component 
analyses for such species as H$_2$, \ion{Cl}{2}, \ion{Fe}{2}, and CO.
The generally similar properties seen for the individual LV components in
the higher resolution optical and STIS spectra, however, suggest that, 
{\it in this particular case}, reasonably accurate total line-of-sight column 
densities may be obtained from the {\it FUSE} spectra by using either 
single-component fits or else multi-component fits in which the relative 
LV component column densities (as well as the $b$-values and relative 
velocities) are all held fixed. This approximation is well-supported by the fact that no inconsistencies were found between the one-component COG, 5-component FIT and AOD results derived for the LV gas species detected in the STIS data, as illustrated by the analysis of neutral carbon (see Fig. \ref{carb} and \S\,4.3). [That was not the case for HD 73882 (Snow et 
al. 2000), where the individual components differed more --- both in 
overall strength and in ratios such as $N$(\ion{K}{1})/$N$(CH).]  
In light of these results, the following study of the physical properties of the cold interstellar gas toward HD\,192639 will be based on the total LV column densities.

\subsection{Molecular lines}


The {\it FUSE} wavelength range contains numerous \Htwo\, lines from the Werner and Lyman bands. Because of the large column densities, the $J=$ 0 to 2 rotational transitions are severely blended with each other, so that their column densities can only be derived using profile fitting techniques. On the other hand, the $J=$ 3 to 5 rotational transitions can be analyzed either via profile fitting or via curve of growth techniques. \citet{Rach01b} performed a multi-component curve of growth analysis of the \Htwo\, lines as part of the \Htwo\,translucent cloud survey. We performed a profile fitting analysis for all \Htwo\, lines, in part to compare results from the two independent methods toward HD\,192639.
The \Htwo\, column densities summarized in Table \ref{low} arose from a 
single-component Voigt profile fitting of the lines  from the $J=$ 0 to $J=$ 5 rotational levels in the {\it FUSE} bandpass. All lines were simultaneously fitted in the 1043-1055, 1073-1083 and 1088-1098 \AA\ windows present in the LiF1A, LiF2A and SiC1A segments (see Fig. \ref{hdeux}, segment LiF1A). As for the atomic species, consistency was found between the COG results from \citet{Rach01b} and our profile fits. These analyses showed that the $J=$ 0 to 2 lines are damped and that the $J=$ 5 lines are optically thin whereas the $J=$ 3 and 4 lines show various degrees of saturation. The best fit (minimized $\chi^2$) led to $b = $(7.2 $\pm$ 0.3) \kms\, (constrained by the $J=$ 3 and 4 lines) using a line spread function of 9 pixels for the {\it FUSE} data. For a check of consistency, we also performed a 5-component fit of the $J=$ 0 to 2 levels using the \ion{K}{1} structure (see \S\, 4.1) and, again, found no discrepancy between the total column densities derived from the 5-component fit and those obtained with the single-component profile fit. The excitation diagram (N($J$)/g$_{J}$ versus E$_{J}$) led to a kinetic temperature derived from the $J=$ 0 and $J=$ 1 rotational levels of T$_{01}=$ (90 $\pm$ 10) K. The diagram also showed that the $J=$ 2 level also appears to be in thermal equilibrium at T$_{01}$, suggesting that the population of that level is still dominated by collisions with thermal atoms.  This analysis resulted in a total \Htwo\, column density of (20.76 $\pm$ 0.09) dex and a total molecular fraction of $f\equiv$ 2N(\Htwo)/[N(\ion{H}{1}) +2\,N(\Htwo)]$=$ (0.36 $\pm$ 0.10) in the LV gas. We used the \ion{H}{1} column density measurement of (21.32 $\pm$ 0.12) dex from \citet{Diplas94}. The higher $J$ levels are not in thermal equilibrium at T$_{01}$ indicating that those transitions are, most probably, predominantly populated by UV pumping \citep{black}. One might expect a broader velocity distribution for those higher levels as indicated by the larger $b$-value found for \Htwo\, (7 \kms\, from $J=$ 3 and 4 $vs$ 4 \kms\, for \ion{C}{1}; see \S\, 4.3) if the excited \Htwo\, arises from less shielded parts of the clouds. Comparable $b$-values were found and discussed by \citet{Jenkins97} in the case of $\zeta$ Ori.\\

Carbon monoxide was detected in the {\it FUSE} and STIS datasets at 1076, 1087 and 1344 \AA. The detailed discussion of the molecular content toward HD\,192639 (including the study of the higher $J$ levels of molecular hydrogen) is addressed in a paper in preparation which compares HD\,192639 to another sightline and discusses the variability of the CO/\Htwo\, ratio (Sonnentrucker et al. 2002 in preparation). We will therefore only mention the main results of the CO analysis toward HD\,192639 here. The detailed profile analysis of the 1076 and 1344 \AA\ bands showed that the latter is optically thin and the former is slightly saturated. The 1087 \AA\ band was not used due to its blend with the 1088 \AA\ \ion{Cl}{1} line (see \S\, 4.6). An estimate of its equivalent width, is, however, given in Table 1. The optically thin 1344 \AA\ band is very weak but shows the component groups seen in \ion{K}{1}, \ion{C}{1} or \ion{S}{1} characteristic of the cold neutral gas. The analysis of the rotational transitions indicates that the band is dominated by the R(0) transition. A small contribution of the $J=$ 1 transitions is however suggested by the 5-component fit based on the \ion{K}{1} structure (see \S\, 4.1 and Table 1). Again, the low resolution of the {\it FUSE} data and the extreme weakness of the 1344 \AA\ band did not allow the study of the CO distribution on a component-by-component basis. Table 3, hence, contains the total line-of-sight CO column density derived from the $J=$ 0 and 1 transitions using the 1076 and 1344 \AA\ bands. We adopted the wavelengths and $f$-values from \citet{morton94} for the 1344 \AA\ band and \citet{federman} for the 1076 \AA\ band.

\subsection{Local hydrogen density: {\bf $n_{\rm H}$}}

Absorption from the ground and excited fine-structure states of \ion{C}{1} is only detected in the LVCs. 
The large sample of lines (24 in all) available in the combined {\it FUSE} and STIS data allowed us to use three of the analysis methods mentioned above and thus to derive accurate column densities for \ion{C}{1}, \ion{C}{1}*, and \ion{C}{1}** as follows:
\begin{itemize}
\item Because the two trace species \ion{K}{1} and \ion{C}{1} should largely coexist, we used the component structure derived from the high-resolution optical spectra of \ion{K}{1} to fit the lines of \ion{C}{1} and excited states seen in the STIS spectra. As noted above, for $T$ $\sim$ 100 K the component $b$-values (for both species) are dominated by turbulence. We, hence, adopted the $b$-values of \ion{K}{1} as fixed parameters. The free parameters in the \ion{C}{1}, \ion{C}{1*} and \ion{C}{1**} fits were the 5 individual component column densities and the continuum. The component-by-component results are summarized in Table 2. This fitting procedure was also used on the {\it FUSE} \ion{C}{1} lines. The total column density derived with the {\it FUSE} data was consistent with the STIS results  within 2$\sigma$. The carbon excited states in the {\it FUSE} data could not be fitted owing to the lower resolution and lower S/N of the data as well as strong blends with other lines.
\item  The equivalent widths of the \ion{C}{1}, \ion{C}{1*} and \ion{C}{1**} lines were measured (Table \ref{ewlow}) and fitted to separate curves of growth (Fig. \ref{carb}). The column densities, $b$-values and respective 1$\sigma$ errors for each species are reported in each panel. The total column densities derived from the curves of growth are consistent 
with those derived from the detailed profile fits within 1$\sigma$.  
For the lines we measured, single-component curves of growth with effective 
$b$-values of 4 km s$^{-1}$ provide reasonable representations of the more 
complex reality; they are consistent with curves generated using the \ion{K}{1} and \ion{Na}{1} component structures. We note that use of the $f$-values suggested by Jenkins \& Tripp (2001) would lead to curves of growth with effective $b$ $\ga$ 5 km s$^{-1}$ and significantly lower column densities.
\item We also integrated the apparent optical depth of the weak \ion{C}{1} (1270.14 \AA), \ion{C}{1*} (1276.75 \AA) and \ion{C}{1**} (1261.43 \AA) lines over the velocity range $-$30 to 0 \kms\, assuming unresolved saturated structures were not present. This last set of results was also in agreement with the fit and COG results within 1$\sigma$.
\end{itemize}
As mentioned in \S\,4.1, the 3 methods gave consistent results within errors which indicates that a single-component analysis provides a fair approximation of the more complicated high-resolution situation, {\it in the particular case of HD\,192639}. From our measurements, the total fractional abundances of carbon in the excited states are: f$_{1}$ $\equiv$ $N$(\ion{C}{1*})/$N(\rm{C_{total}})=$ 0.134 and f$_{2} \equiv N$(\ion{C}{1**})/ $N(\rm{C_{total}})=$ 0.020, leading to a pressure estimate $\log (P/k) = \log (n_{\rm H}\,T)$ between 3.1 and 3.3 for the LV gas \citep{Jenkins79}, if the gas is bathed in the average interstellar radiation field. Assuming a kinetic temperature for carbon of 100 K, in agreement with that of \Htwo\,(T$_{01}$), the number density of neutral hydrogen atoms (averaged over the 5 components) $n_{\rm H}$ is estimated to range from 12 to 20 cm$^{-3}$ in the LVCs. This range implies that the \ion{H}{1}-containing gas has a total thickness between $\sim$ 34 and 57 pc. \\

The 1355 \AA\ \ion{O}{1} line and the excited fine-structure states \ion{O}{1*} (1304 \AA) and \ion{O}{1**} (1306 \AA) were also detected in the LV gas (Fig. \ref{velb}). We derived the \ion{O}{1} column density using the methods described earlier for neutral carbon. The \ion{O}{1*} and \ion{O}{1**}  lines are assumed optically thin allowing derivation of their respective column densities from the equivalent width measurements. A comparison of the profiles (Fig. \ref{profile}), however,
suggests that the column density distribution in \ion{O}{1*} and \ion{O}{1**} is different from that in \ion{O}{1} or \ion{C}{1}. In order to quantify these profile variations, we performed tests by fitting the \ion{O}{1*} and \ion{O}{1**} profiles using \ion{O}{1} and \ion{C}{1} as the baseline for the initial 5-component structures. The $b$-values were fixed to those of \ion{K}{1} for each component (see \S\, 4.1). The total column densities are consistent with the column densities derived from the equivalent width analysis (see Tables 2 and 3). We, therefore, reported the weighted mean of the column densities for \ion{O}{1}, \ion{O}{1*} and \ion{O}{1**} in Table \ref{low}.
We adopted the latter total column densities, and used the \citet{Keenan88} calculations of the fine-structure population ratios, to estimate the neutral hydrogen volume density in the LVCs where \ion{O}{1} is detected. We find population ratios of \mbox{$\log$[f$_{1}$ $\equiv$ $N$(\ion{O}{1*})/$N(\rm{O_{total}})$]$= -$5.25} and $\log$ [f$_{2}$ $\equiv$ $N$(\ion{O}{1**})/$N(\rm{O_{total}})$]$= -$5.43. The best match with \citet{Keenan88} calculations occurs for T=100 K and \mbox{11 $\le$ $n_{\rm H}$ $\le$ 26 cm$^{-3}$}, which is consistent with the volume density derived from the \ion{C}{1} analysis. The \ion{C}{1} and \ion{O}{1} volume densities are of the order of those found in diffuse clouds, indicating that the LVCs are most probably dominated by diffuse cloud gas. The component-by-component analysis (Table 2) suggests that potentially slightly denser material is concentrated in component 3 in the LVCs. However, the weakness of the \ion{O}{1**} line, the poor S/N in the line and the uncertainties linked to the individual column density determinations (see \S\, 4.1) prevent us from discerning whether the \ion{O}{1**} line profile resembles that of \ion{O}{1*} or is different. New high-resolution and high S/N observations of those atomic lines are warranted to confirm the profiles presented in this study as well as the potentially denser neutral hydrogen clumping.

\subsection{Depletion}

Except for \ion{C}{1}, \ion{Na}{1}, \ion{Mg}{1}, \ion{S}{1} and \ion{K}{1}, the species detected in the neutral gas are in their dominant charge state, allowing us to derive their  depletions. We used the solar abundances from \citet{anders89} and \citet{greve93} to estimate 
the depletions of \ion{Mg}{0}, \ion{Cl}{0}, \ion{Mn}{0}, \ion{Fe}{0}, \ion{Ni}{0}, \ion{Cu}{0} and \ion{Ge}{0}. The solar abundance for \ion{O}{0} was taken from \citet{hol01}. The strong saturation of \ion{S}{2} ($\lambda$1250), \ion{Si}{2} ($\lambda$1193) and \ion{C}{2} ($\lambda$1334) prohibits us from deriving any information on the depletion of these species. We will, however, assume abundances to be solar for sulfur and adopt the fairly constant interstellar carbon depletion of $-$0.4 dex from solar \citep{sophia97} in the following discussion on electron densities. Table 2 and Fig. 4 show that all refractory elements seen in STIS exhibit very similar profiles suggesting that abundances and depletions are very similar in each component. Hence, as in the case of the trace neutrals, the total LV gas abundances and depletions can and will be considered. Table \ref{deple1} summarizes the abundances relative to hydrogen and the depletions for the LV gas (over the 5 components). Figure \ref{deple} compares the estimated depletions with the average depletions for the typical ``cold'' and ``warm'' clouds adopted by Welty et al. (1999; see references therein). If the  HD\,192639 line of sight traversed a translucent cloud, one would expect the depletions to be of the order of, or more severe, than those in the cold diffuse clouds \citep{snow75}. However, the LV gas does not show enhanced depletions of the species forming dust grains (\ion{Mg}{2}, \ion{Mn}{2}, \ion{Fe}{2}, \ion{Ni}{2} or \ion{Cu}{2}) but rather shows depletions intermediate between cold and warm diffuse clouds. Therefore, even though the line of sight toward HD\,192639 has been considered as a candidate for containing a translucent cloud on the basis of its reddening, this depletion study supports the diffuse-cloud dominated nature of the LV gas, already suggested by the complexity of the velocity structure shown in the trace neutral species and the volume density derived from the carbon analysis. This line of sight should therefore be called ``translucent line of sight'' rather than ``translucent cloud'' \citep{Rach01b}.

\subsection{Electron Density and Fractional Ionization}

We used neutral/first ion pairs for five elements (C, Na, Mg, S and K)  and the assumption of photoionization equilibrium to estimate the electron density in the LVCs (Table \ref{dens}). Except for Mg, we do not have measured column densities for the dominant first ions, and so we estimated values based on assumed depletions: 0.0 dex for S \citep{fitz93}, $-$0.4 dex for C (Sofia et al. 1997), $-$0.6 dex for K, and $-$0.7 dex for Na. The values for K and Na were estimated based on considerations of photoionization equilibrium (Welty \& Hobbs 2001), and may be too severe if other processes (e.g., charge exchange with large molecules; Weingartner \& Draine 2001) affect the ionization balance of the heavy elements.  
The photoionization rates were calculated using the average IS radiation fields of Draine (1978) and Witt \& Johnson (1973); the recombination rates were 
calculated for $T =$ 100 K (P\'{e}quignot \& Aldrovandi 1986).  
The electron densities inferred from C, Na, Mg, and K are fairly similar 
(0.11 $\pm$ 0.02) cm$^{-3}$; the much lower value from S has been noted in other 
sightlines (Welty et al. 1999; Welty et al. 2002a). \\

On the basis of the carbon analysis and the previous depletion discussion, we adopted a mean electron density of (0.11 $\pm$ 0.02) cm$^{-3}$ and mean neutral hydrogen density of (16 $\pm$ 3) cm$^{-3}$ indicating an average ionization fraction $n_{e}$/$n_{\rm{H}}$ $\sim$ 0.007, which is higher than that expected from the carbon ionization only (1.4$\times$10$^{-4}$). Such high fractional ionizations have been seen in other sightlines (e.g.,
23 Ori; Welty et al. 1999; Welty et al. 2002a), and may be due (at least in part) to overestimation
of $n_e$ if processes such as charge exchange with large molecules are
significant (Lepp et al. 1988; Weingartner \& Draine 2001). On the other hand, the presence of \ion{N}{2} and maybe \ion{N}{2*} in the LVCs also suggests there is some ionized gas amid the neutral gas which could partly explain the high ionization fraction toward HD\,192639.  \\

\subsection{Chlorine chemistry}

In regions where H$_2$ is optically thick, \ion{Cl}{2} reacts rapidly with 
\Htwo\, to form
HCl$^{+}$, which in turn (via several routes) leads to \ion{Cl}{1} and 
\ion{H}{1}.  
Because the conversion of \ion{Cl}{2} to \ion{Cl}{1} is faster (rate constant k 
= 7 
$\times$ 10$^{-10}$ cm$^3$s$^{-1}$) than the photoionization of \ion{Cl}{1}
($\Gamma$ = 2 $\times$ 10$^{-12}$ s$^{-1}$), Cl is primarily neutral in regions 
where H$_2$ is abundant, but otherwise is primarily ionized (Jura 1974; Jura \& 
York 1978). This behavior is unique to Cl.  
Jura \& York proposed a simple cloud model consisting of two zones:  1) a region
where H$_2$ is optically thick, with H$_2$, \ion{Cl}{1}, and some \ion{H}{1}, 
and 2) a 
region where H$_2$ is optically thin, dominated by \ion{H}{1} and \ion{Cl}{2}.
In principle, each cloud along a line of sight could have these two zones.
Equation A7 in Jura \& York then implies that 
$N_1$(\ion{H}{1}) = $N_{\rm H}$ [ $f$(\ion{Cl}{1}) $-$ $f$(H$_2$)],
where $N_1$(\ion{H}{1}) is the column density of \ion{H}{1} in the optically 
thick region, $N_{\rm H}$ is the total hydrogen (atomic plus molecular) column 
density for the line of sight and $f$(\ion{Cl}{1}) and $f$(H$_2$) are the 
fractions of Cl in neutral form and of hydrogen in molecular form, respectively, 
averaged over the line of sight.\\

Chlorine lines were detected in the LVCs at 1004, 1088, 1097 (\ion{Cl}{1}), and 
1071 \AA\ (\ion{Cl}{2}) with {\it FUSE} and at 1347 \AA\ (\ion{Cl}{1}) with 
STIS; several other weak \ion{Cl}{1} lines may be present in the {\it FUSE} 
spectrum.
The \ion{Cl}{2} column density was estimated using the equivalent width and 
with the AOD method, assuming the line is optically thin.
Both methods gave results consistent with $N$(\ion{Cl}{2}) $\sim$ 1.7 $\pm$ 0.4$\times$10$^{14}$ cm$^{-2}$.
All three of the \ion{Cl}{1} lines are somewhat saturated; the line at 
1088.059 \AA\ is slightly blended with a CO line at 1087.868 \AA\ (see Fig.~2).
Unfortunately, only the \ion{Cl}{1} lines at 1088, 1097, and 1347 \AA\ have 
fairly reliable, experimentally determined $f$-values (Schectman et al. 1993).
Application of the AOD method to the $\lambda$1347 line implies 
$N$(\ion{Cl}{1}) $>$ 1.1$\times$10$^{14}$ cm$^{-2}$.
Fits to the {\it FUSE} lines, using the component structure derived from the higher resolution \ion{K}{1} and \ion{Na}{1} data, with various constraints on the relative column densities of the five strongest components, suggest that $N$(\ion{Cl}{1}) $\sim$ 1.1-1.5 $\times$ 10$^{14}$ cm$^{-2}$. Fits to the $\lambda$1347 line, using the same component structure and constraints on the relative column densities of the five strongest components, suggest that $N$(\ion{Cl}{1}) $\sim$ 2--3 $\times$ 10$^{14}$ cm$^{-2}$. Those combinations of component structure and total column density yield W(1088) $\sim$ 50 m\AA\ (vs. the observed 50 $\pm$ 15 m\AA), but appear to somewhat overestimate the strength of the weaker {\it FUSE} $\lambda$1097 line. In the light of the AOD and FIT results, we will adopt $N$(\ion{Cl}{1}) $\sim$ 2.0$^{+1.0}_{-0.5}$ $\times$10$^{14}$ cm$^{-2}$. 

The adopted column densities imply $f$(\ion{Cl}{1}) = 0.54 --- similar to
the values found in a survey of less heavily reddened sightlines, for 
comparable $f$(H$_2$) (Jenkins et al. 1986).
With $f$(H$_2$) = 0.36, we find $N_1$(\ion{H}{1}) = 5.8$\times$10$^{20}$ 
cm$^{-2}$ --- so that about 28\% of the \ion{H}{1} is in zone(s) 1, 
where H$_2$ is abundant and optically thick.
The $f$(H$_2$) in zone(s) 1 would then be about 0.66 --- i.e., higher than the 
average value 0.36 for the whole line of sight, but somewhat lower than the 
values expected for translucent clouds.
If $N$(\ion{Cl}{1}) were as low as 1.5$\times$10$^{14}$ cm$^{-2}$, then the
$f$(H$_2$) in zone(s) 1 could be as high as 0.75 --- but note the effects of
possible differences in depletion discussed in the next paragraph.
Some of the \ion{H}{1} in zone 2, perhaps of order 2--3$\times$10$^{20}$ 
cm$^{-2}$, could be associated with the weak, outlying components seen in 
\ion{Na}{1} (Welty \& Hobbs 2001).

While the model of Jura \& York (1978) assumes that the depletion of 
Cl is the same in both zones, that assumption may not be valid, as Cl appears
to be slightly more severely depleted in sightlines with higher mean $n_{\rm H}$
(e.g., Jenkins et al. 1986). 
If that assumption is dropped, equation A7 in Jura \& York can be generalized 
to $f_1$ = $f$(\ion{Cl}{1}) [1 + $D$ $f_2$] $-$ $f$(H$_2$), where
$f_i$ = $N_i$(\ion{H}{1}) / $N_{\rm H}$ is the fraction of the total hydrogen 
as \ion{H}{1} in zone $i$ [i.e., $f_1$ + $f_2$ + $f$(H$_2$) = 1].
The quantity $D$ = ($d_2$ $-$ $d_1$) / $d_1$, where $d_1$ and $d_2$ are the 
depletions of Cl in the two zones.
Further manipulation of the above equation yields 
$f_2$ = [1 $-$ $f$(\ion{Cl}{1})] / [1 + $D$ $f$(\ion{Cl}{1})].
The observed $f$(\ion{Cl}{1}) and $f$(H$_2$), together with the constraints
0 $\le$ $f_i$ $\le$ 1 $-$ $f$(H$_2$), then allow us to place limits on the 
relative depletions of Cl in the two zones.
As the overall depletion of Cl toward HD 192639 is very mild, however, in 
this case the depletions in zones 1 and 2 are likely to be fairly similar.
Note that for given values of the total sightline column densities, more 
severe depletions of Cl in zone(s) 1 [versus those in zone(s) 2] would imply 
more \ion{H}{1} in zone(s) 1 --- and thus a {\it lower} $f$(H$_2$) there than 
if the depletions were equal in the two zones.\\

In principle, such analyses enable the \ion{H}{1} column densities, the
Cl depletions, and the molecular fractions to be estimated separately for both 
zones, given only the integrated sightline values.
In particular, we may thus be able to better identify and characterize the 
translucent portions of clouds in a complex line of sight.
Obviously, accurate values for $f$(\ion{Cl}{1}) and $f$(H$_2$) are required;
the former would greatly benefit from better $f$-values for the \ion{Cl}{1}
lines observable with {\it FUSE}.
It will be very interesting to examine these issues for other more reddened 
sightlines. For HD\,192639 also note that, independent of incertainties of $f$-values, $f$(H$_2$)$\sim$1.7 times higher than the integrated line of sight value. Similar analyses may show that the integrated values of $f$(H$_2$) are lower limits to $f$ for the densest clouds along reddened lines of sight.

\section{The High-Velocity Components (HVCs)}
\subsection{Previous Observations}

Three members of Cygnus OB1 association were observed by \citet{phi84} with {\it IUE}. Their study showed the existence of high-velocity components in two lines of sight including HD\,192639. The average velocity of that gas is consistent with the velocity of the components called HVCs in this paper. Their study showed that the HVCs are seen primarily in highly ionized species such as \ion{C}{4}, \ion{Si}{4} (see Fig. \ref{veld}), \ion{S}{3} or \ion{Al}{3}, but lower ionization states from \ion{Fe}{2} and \ion{Mg}{2} are also detected. \citet{nic93} combined {\it IUE} data for 22 stars spanning the OB1 and OB3 associations together with {\it IRAS} maps of the entire region in an attempt to identify the HVCs with shells seen in IR emission. Their analysis concluded that the UV absorption lines detected with {\it IUE} at high-velocity were pervasive throughout the Cygnus associations and appeared to trace a superbubble encompassing OB1 and OB3. They also found the gas to have more high-velocity structures toward Cyg OB1 than Cyg OB3, indicative of gas specifically associated with Cyg OB1.  

\subsection{Ionization}

In the {\it FUSE} and STIS data, the HVCs are seen clearly in
lines from various singly and doubly ionized species (e.g., \ion{C}{2}, \ion{N }{2}, \ion{Si}{2}, \ion{Si}{3}, \ion{S}{2}, \ion{S}{3}); absorption from \ion{N}{1} is also present (Fig. 3). The predominance of ionized gas (see below) and the relatively high velocities ($-$80 \kms) suggest that the 
HVCs probably originated from the interaction of a shock with gas local to the Cyg OB1 association. Part of that shocked gas has cooled and recombined to yield the nitrogen detected in the STIS data. It is also possible that the gas is photoionized by the nearby stars rather than shock ionized, but the presence of the high velocity components leads us to favor the shock hypothesis. The small structures seen in the \ion{Si}{2*} and \ion{S}{2} lines at between $-$70 and $-$80 \kms\, (Fig. 4) suggest the existence of small scale clumps in the swept-up shell, as suggested by \citet{nic93}. \\

The origin of the shocked gas was discussed  by \citet{phi84} based on energy considerations. They concluded that the high-velocity gas could originate either from a supernova or from the combined stellar winds from the hot stars present in Cyg OB1, if the ambient interstellar density is below 0.6 cm$^{-3}$ in order to satisfy the measured stellar wind power produced by the stars in the association \citep{abbott}. We will shortly revisit that question in \S\,5.4.\\

The extended velocity distribution ($-$70 to $-$30 \kms) of ions typically tracing \ion{H}{2} region
gas such as \ion{N}{2} and \ion{N}{2*} or \ion{Si}{2*} suggest that large \ion{H}{2} regions are present over the entire velocity range. Such \ion{H}{2} structures were already mentioned by \citet{loz98a} and \citet{loz98b}. 

\subsection{Electron density and abundances}

The column densities of various species detected in the 
HVCs, determined via the AOD method over the interval $-$100 to $-$50 \kms, are
listed in Table 6.  The $b$-values of the components seen in the \ion{Si}{2*} 1264.7 \AA\ line are of order 2.5 \kms, indicating a maximum temperature of order 10,000 K for these predominantly ionized HVCs.  In such warm, ionized gas, the excited fine structure level of \ion{Si}{2} is typically populated 
primarily via collisions with electrons.  The observed ratio N(\ion{Si}{2*})/N(\ion{Si}{2}) then implies $n_e$ = 1.6$^{+0.7}_{-0.3}$ cm$^{-3}$ \citep{york79}. 

The column densities listed in Table 6 can be used to estimate N(\ion{H}{1}) and N(\ion{H}{2}) in the HV gas. \ion{N}{1}, with ionization potential
14.5 eV, traces predominantly neutral gas; \ion{N}{2} (and \ion{N}{3}) trace ionized gas. 
While the \ion{N}{2} 1083 \AA\ line is strongly saturated, we may use N(\ion{N}{2*}) and the $n_e$
derived above to estimate N(\ion{N}{2}) $\sim$ 6.2$\times$10$^{15}$ cm$^{-2}$ \citep{york79}.
Using the solar abundance of $-$4.03 dex (Grevesse \& Noels 1993) for N, we obtain
N(\ion{H}{1}) $\sim$ 6.3$\times$10$^{16}$ cm$^{-2}$ and N(\ion{H}{2}) $\ge$ 6.9$\times$10$^{19}$ cm$^{-2}$ --- so the HV gas has a
neutral fraction of at most 0.1\%. Note, however, that \ion{N}{1} could be 
underabundant due to its ionization when hydrogen is mostly ionized (Jenkins et al. 2000). In that case our estimate of the neutral hydrogen based on neutral 
nitrogen may only be a lower limit. The ratio N(\ion{H}{2})/$n_e$ implies a total thickness of about 14 pc for the HV gas.  If sulfur is not depleted, then in principle we could estimate N(H) from N(\ion{S}{2}) + N(\ion{S}{3}).  Since the \ion{S}{3} 1190 \AA\ line is saturated and blended with LV \ion{C}{1} absorption, we infer N(H) $>$ 6$\times$10$^{19}$ cm$^{-2}$ from N(\ion{S}{2}) --- consistent with the value obtained from nitrogen.
Because of the relatively high N(H) in the HV components, such lines as \ion{Si}{3} $\lambda$1206, \ion{Al}{2} $\lambda$1670, and \ion{Fe}{3} $\lambda$1122 are strongly saturated making estimates of possible depletions of Si, Al, and Fe in the HV gas very difficult.
We, however, report a lower limit on the column density of \ion{C}{3} estimated by applying the AOD method in the velocity range considered throughout this analysis ([$-$100,$-$50] \kms). \\

While HV gas at similar velocities has also been observed 
toward a number of stars in Orion (Cowie, Songaila, \& York 1979; Welty et al.
1999 and in prep.), much more HV gas is present toward HD\,192639.  For species
whose column densities have been measured or inferred in the LV and HV regions, the values toward HD\,192639 are typically higher by a factor of order 100.  The higher column densities have enabled the detections of both neutral species
such as \ion{N}{1} (and also \ion{Na}{1}) and more highly ionized species such as \ion{Si}{4} and \ion{C}{4}. The strong saturation of lines from a number of other species toward HD\,192639, however, makes detailed comparisons of ion ratios difficult.  Interestingly, the overall N(\ion{H}{1})/N(\ion{H}{2}) and $n_e$ are fairly similar in the two regions, despite the large differences in N(\ion{H}{2}). \\

If the HV gas comes from a radiative shock, as conjectured earlier, it would be interesting to estimate the amount of energy injected into the ISM in order to produce a total hydrogen column density of $\sim$ 7$\times$10$^{19}$ cm$^{-2}$ at a projected velocity of $-$80 \kms. To do so, one however needs some knowledge about the geometry of the shocked region. We discuss this in the next section.

\subsection{The Energy Balance}

As mentioned in \S\, 5.2, \citet{abbott} measured the stellar wind power produced by all the stars in the Cyg OB1 association to be about $\sim$ 1.2$\times$10$^{38}$ erg s$^{-1}$. If those stars have been sustaining those winds for 4$\times$10$^{5}$ years, the energy input from these winds in the ISM is about 1.5$\times$10$^{51}$ erg.
\citet{nic93} noted weak emissions centered on the Cyg OB1 association (2$^{\circ}$$\times$4.5$^{\circ}$ in extent) in the IRAS data. In addition, the {\it IUE} data, which they re-analyzed, indicated traces of a multi-association superbubble of, at least, 4$^{\circ}$$\times$6$^{\circ}$ in extent. If the HVCs originate in the first or the second bubble, the linear dimensions of the bubble would be (62$\times$140)pc or (126$\times$190)pc, respectively. For the purpose of this exercise, we assume the 2 bubbles are spherical and have respective linear radii of 31 pc and 63 pc. If we further assume that the total hydrogen in the HVCs is homogeneously distributed in the 31 pc or the 63 pc radius shell, we can estimate the mass contained in the shells using the column density of $\sim$ 7$\times$10$^{19}$ cm$^{-2}$, which we derived from the N and S analyses in the previous section. We find a mass of 6.7$\times$10$^{3}$ M$_{\odot}$ and 2.8$\times$10$^{4}$ M$_{\odot}$ for the shell with a radius of 31pc and 63pc, respectively. The corresponding kinetic energies for a projected velocity of 80 \kms\, are 4.2$\times$10$^{50}$ erg and
1.8$\times$10$^{51}$ erg for the 2$^{\circ}$ and the 4$^{\circ}$ shells respectively. Are the stellar winds in Cyg OB1 capable of producing such energy? How does it compare with supernova energy injection? The model of spherical wind bubble developed by \citet{weaver77}, with the above-mentioned radius, HV gas velocity and stellar wind energy (1.5$\times$10$^{51}$ erg) derived from \citet{abbott}, indicates that the ambient pre-shock density has to be lower than 0.6 cm$^{-3}$. If we assume this value to be our upper limit on the ambient density in that region, the corresponding upper limits on the energy produced by a supernova \citep{chevalier74} are 6.2$\times$10$^{50}$ erg (31pc) and 5.7$\times$10$^{51}$ erg (63pc). The comparison of those estimates with the kinetic energies required to produce the hydrogen shells considered above, indicates that the distinction  between a stellar wind or a supernova origin cannot be made on energy considerations only, under our assumptions. \citet{loz98b} explained the multi-shell structure seen toward Cyg OB1 and OB3, as the result of a 2-step kinematics in which the winds from the OB stars create shells of low-velocity gas with which the winds of WR stars (or supernovae appearing later in the association's history) interact, forming higher-velocity shells. A study of the HV gas ionization structure would be required to probe their conclusions. This study will be presented in a separate work.  

\section{Summary/Perspectives}

We have presented here a survey of the detailed analyses that can be performed when using data from different instruments. We combined the far-UV dataset from {\it FUSE}, {\it HST}/STIS and {\it IUE} archival datasets with high-resolution optical data and focused on the study of the abundances and physical conditions of the low-velocity interstellar gas toward the candidate translucent
cloud star HD\,192639. A brief overview of the high-velocity gas components seen along this line of sight was also given. The high-resolution optical \ion{K}{1} data were used to define the velocity structure in the neutral gas that could not be resolved with the {\it FUSE} data (18 \kms) and could not be entirely resolved with the STIS data (2.75 \kms). We did not use a specific velocity structure model when analyzing the highly-ionized gas. \\

The neutral gas exhibits multiple low-velocity components (LVCs) at velocities between $-$18 and $-$8 \kms (heliocentric). It consists of gas with temperatures and densities typical of cold diffuse clouds mixed with slightly warmer gas. The analysis of the \Htwo\, $J=$ 0 and 1 transitions indicates a temperature of that gas of (90 $\pm$ 10) K. The higher rotational transitions ($J=$ 2) seem to be in thermal equilibrium at T$_{01}$. The trace neutral elements \ion{C}{1} and \ion{S}{1} show profiles very similar to that of \ion{K}{1}, suggesting 
their co-location in the LV gas. Analysis of the \ion{C}{1} fine-structure excitation led to an estimate of the line-of-sight neutral hydrogen density of (16 $\pm$ 3) cm$^{-3}$, typical for cold diffuse clouds.
The detection of \ion{O}{1*} and \ion{O}{1**} at the velocity of the LVCs, suggests densities very similar to those derived from the carbon analysis, ranging from 11 to 26 cm$^{-3}$. 
The depletion study shows that the average properties of the LV gas toward HD\,192639 are close to those exhibited by diffuse clouds rather than by translucent clouds \citep{snow75,snow00,rach01a}. The line of sight seems dominated by multiple diffuse clouds, as already indicated by the velocity structure and supported by the component-by-component analysis of the partially resolved STIS data. Analysis of the overall \ion{Cl}{1}, \ion{Cl}{2}, and f(\Htwo) indicates higher molecular fractions in the regions where \Htwo\, is dominant and confirms the diffuse nature of the cold LV gas. The average electron density, derived from five pairs of elements, is estimated to be (0.11 $\pm$ 0.02) cm$^{-3}$--- leading to an ionization fraction in the LVCs of 0.007, much larger than expected from carbon ionization alone. The LVCs seem therefore to contain some ionized hydrogen in this predominantly neutral material. Alternatively, the high fractional ionization could be partly due to an overestimation of $n_e$ if the ionization equilibrium is affected by other processes besides photoionization and radiative recombination.\\

Our overview study of the high velocity gas shows that the latter is also made of multiple blended components (HVCs) ranging in velocity from $-$60 to $-$80 \kms\, (heliocentric). This gas is predominantly ionized, but also contains a small amount in neutral form at the same velocity, suggesting fast cooling and recombination. Some of the material may be due to a radiative shock. The silicon fine-structure excitation analysis suggests that the H II region has an electron density of the order of 1.6 cm$^{-3}$. A lower limit estimate to the total hydrogen column density of $\sim$ 7 $\times$10$^{19}$ cm$^{-2}$ was derived from the nitrogen and sulfur measurements in the HV gas. The kinetic energy injected into the ISM to generate the measured hydrogen column density at the HV gas velocity can be produced by either stellar winds or supernovae.\\

Complementary {\it HST} data would be useful to get \ion{Mg}{1}, \ion{Fe}{1}, \ion{S}{2}, \ion{Si}{1}, \ion{Si}{2} and \ion{Zn}{2} in order to derive, directly, the electron densities and further study the depletions in the LV gas. Further observations of Cyg OB1 stars located in a 3$^{\circ}$ radius of HD\,192639 would provide unique insights in the small-scale density (and pressure) variations of the ISM as seen toward HD\,192639. SNR shocks show further signatures in the optical in the form of characteristic emission lines such as those from [\ion{O}{3}], [\ion{S}{2}] or [\ion{N}{2}]. Optical observations of HD\,192639 performed offset from the star itself would allow a search for those emission lines and provide a test for the supernova remnant shock origin of the HVCs toward HD\,192639. A study of the \ion{Cl}{1} $f$-values in the {\it FUSE} bandpass is warranted. 



\acknowledgments

This work is based on data obtained for the Guaranteed Time Team by the NASA-CNES-CSA {\it FUSE} mission operated by the Johns Hopkins University. Financial support to U. S. participants has been provided by NASA contract NAS5-32985. D. E. Welty acknowledges support from the NASA LSTA grant NAG5-3228 to the University of Chicago. This work has been done using the profile fitting procedure Owens.f developed by M. Lemoine and the {\it FUSE} French Team. This research has made use of the SIMBAD database, operated at CDS, Strasbourg, France. The authors thank M. Andr\'{e} for fruitful comments on the {\it FUSE} data reduction.

\clearpage


\begin{figure}
\caption{Absorption from \ion{C}{2}, \ion{N}{1} and \ion{O}{1} in the LVCs and the HVCs.\label{velb}}
\plotone{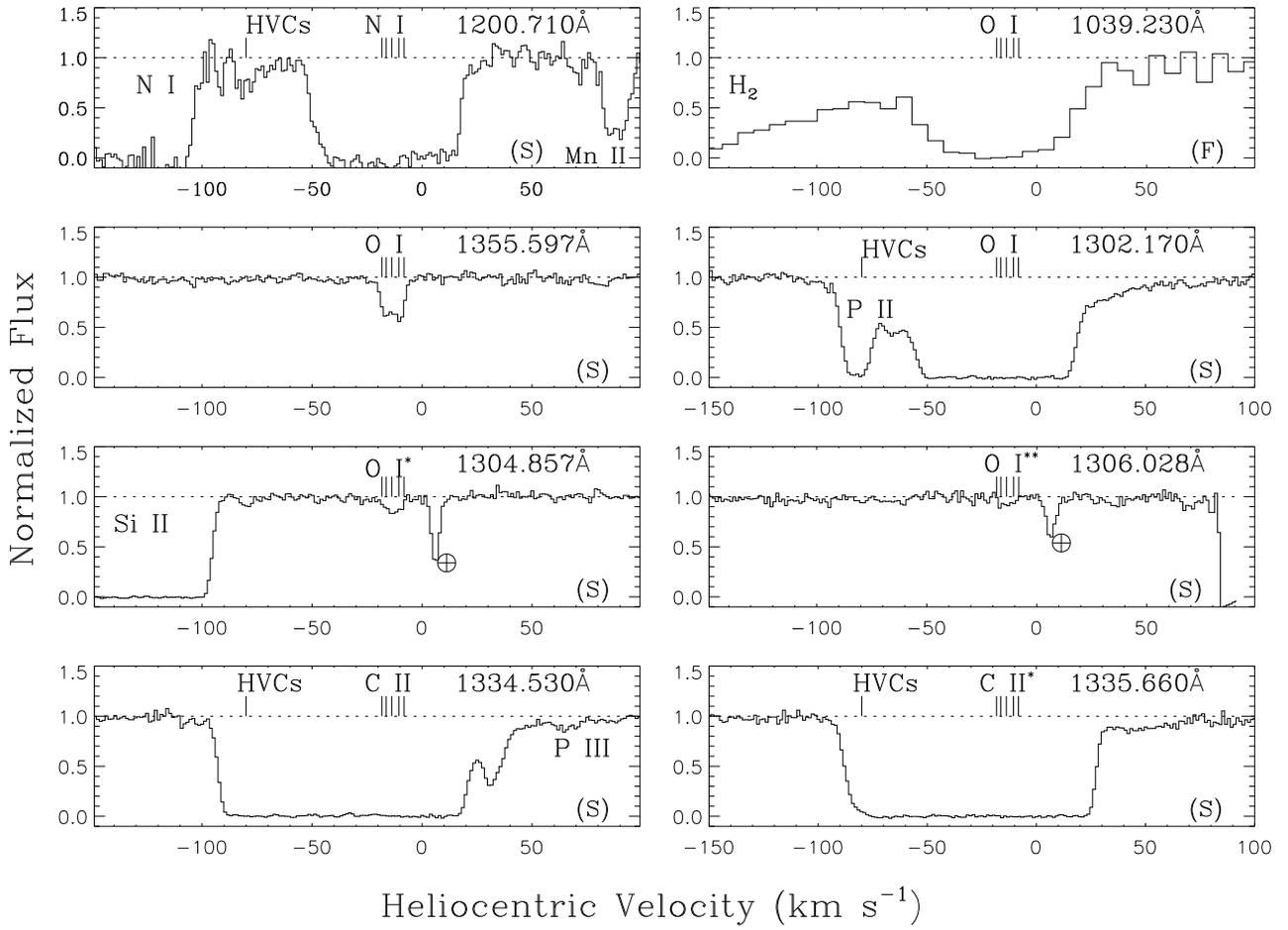}
\end{figure}

\clearpage

\begin{figure}
\caption{Absorption from singly ionized species in the LVCs and the HVCs.\label{velc}}
\plotone{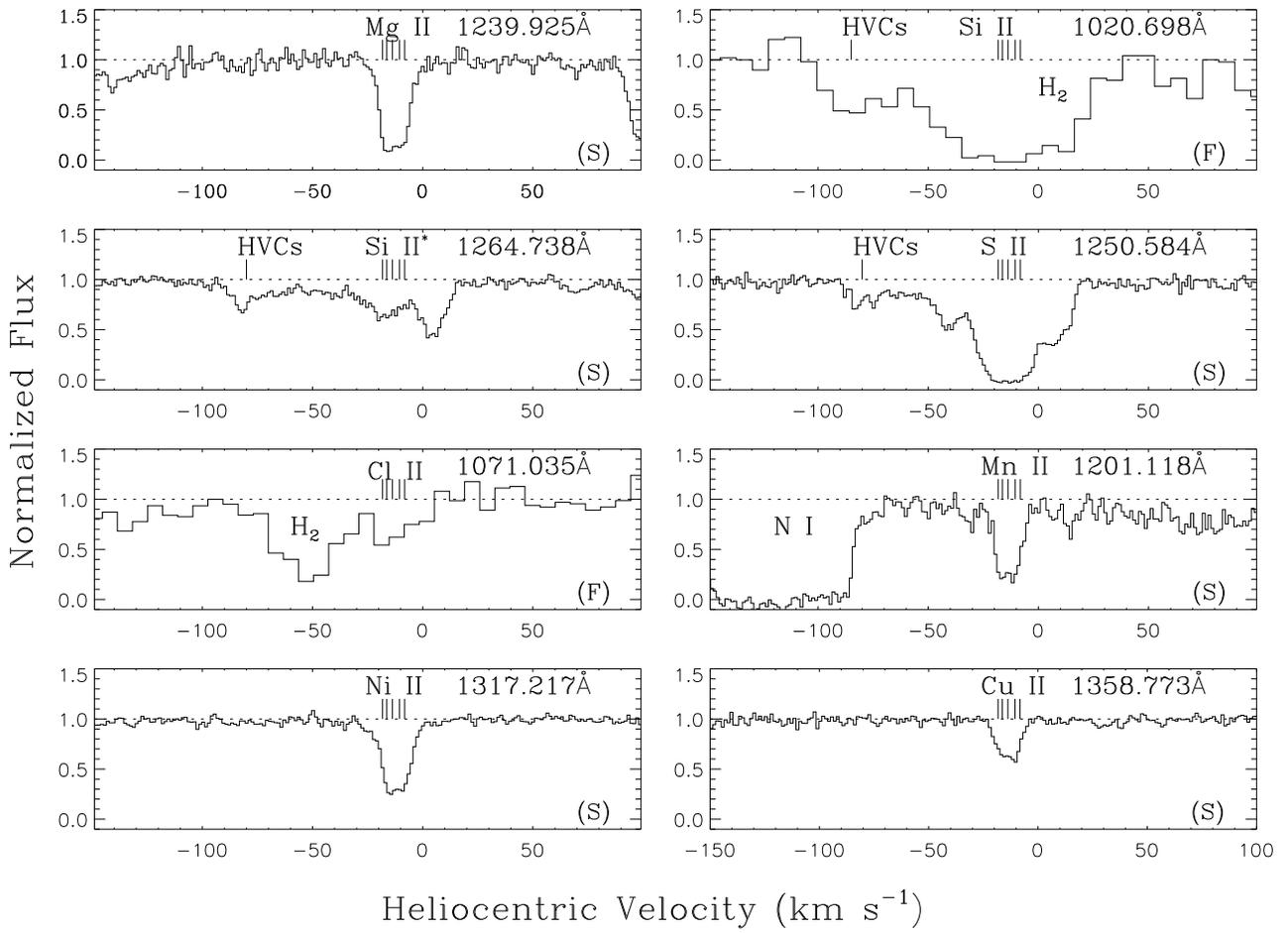}
\end{figure}

\begin{figure}
\caption{Strong lines from ionized species clearly delineate the HV gas.\label{veld}}
\plotone{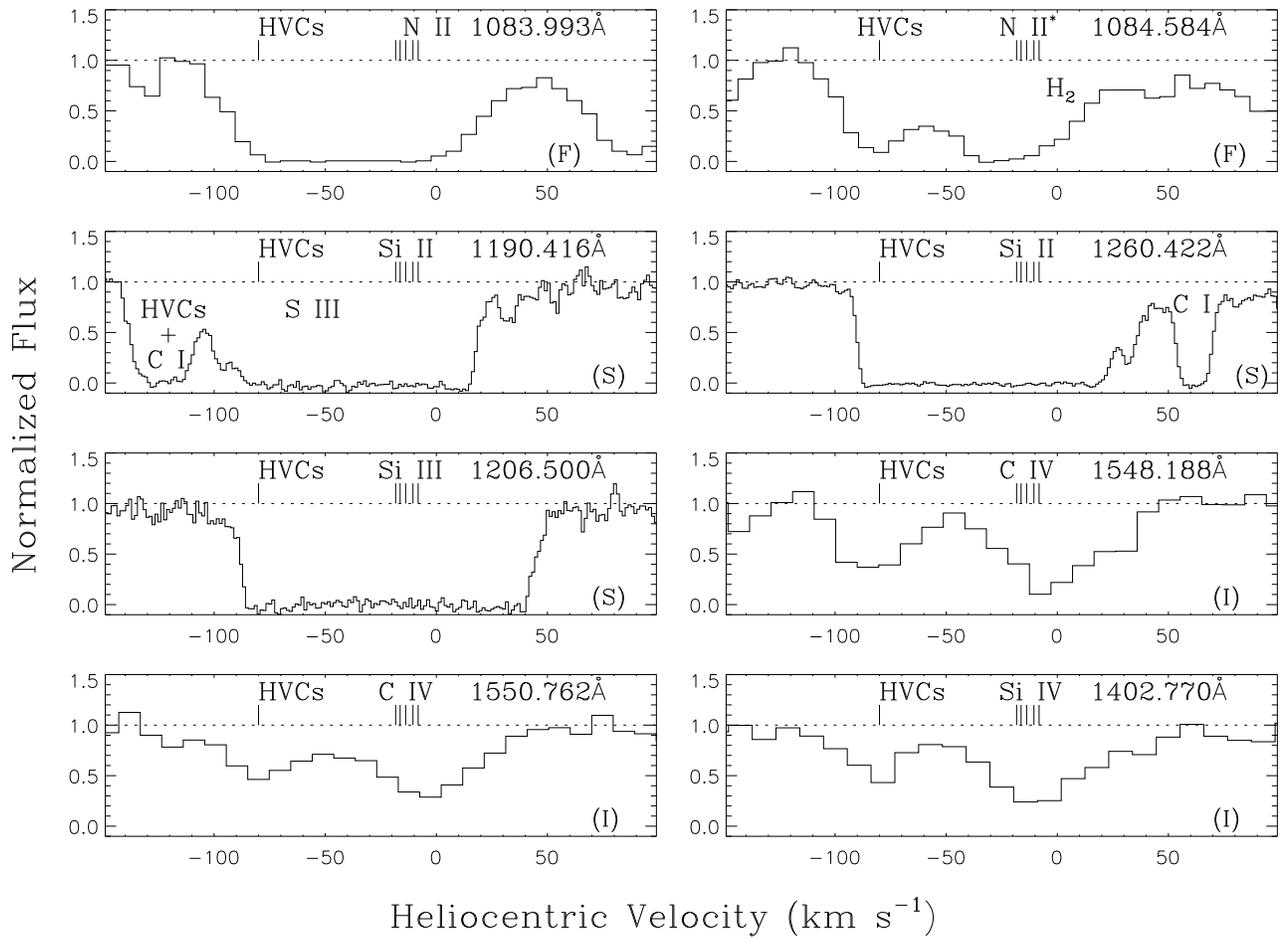}
\end{figure}


\clearpage

\begin{figure}
\caption{Comparison of the \ion{O}{1}, \ion{O}{1*} and \ion{O}{1**} profiles in the LV gas. The '{\bf $\oplus$}' symbol denotes the position of the telluric lines. The tick marks indicate the position of the 5 components forming the LVCs.\label{profile}}
\plotone{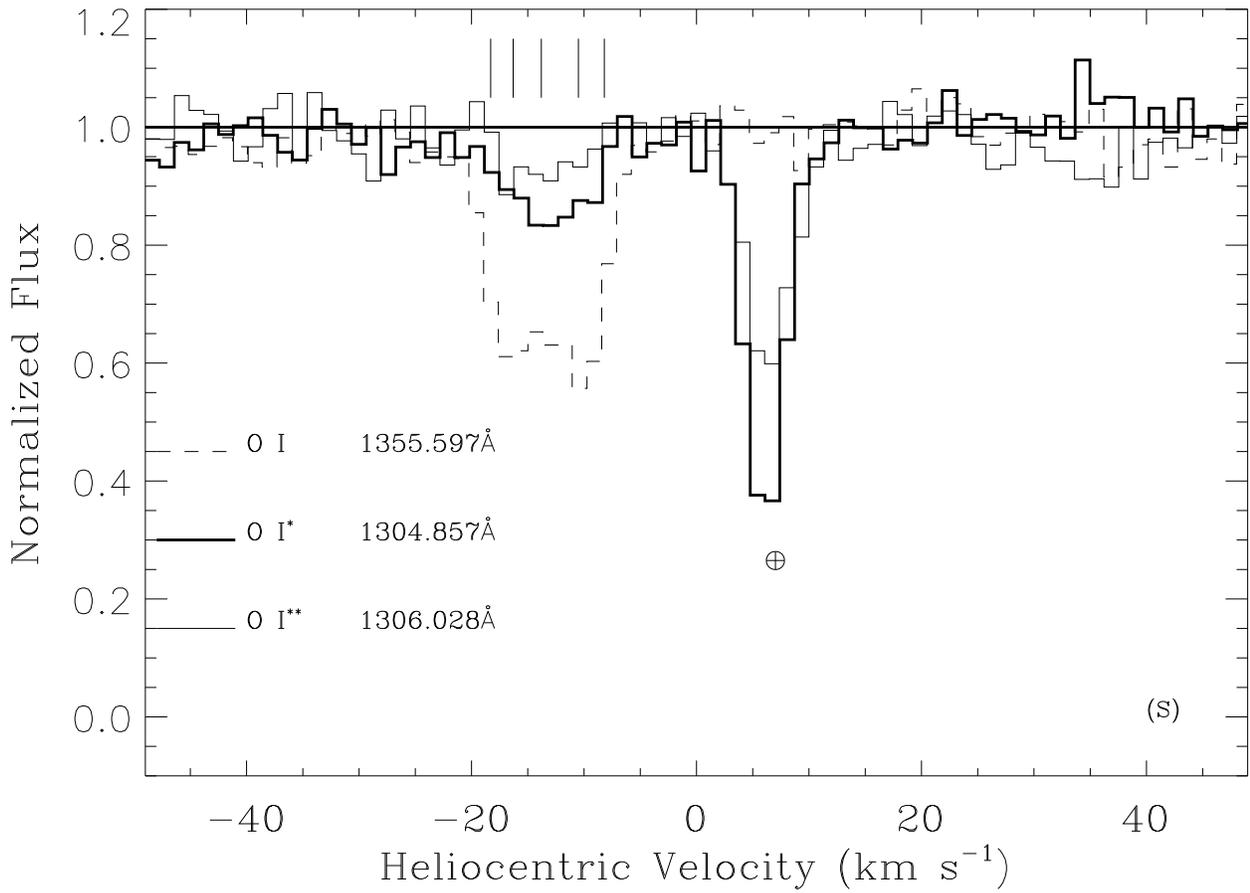}
\end{figure}

\begin{figure}
\caption{Curve of growth for \ion{C}{1}, \ion{C}{1*} and \ion{C}{1**} detected with {\it FUSE} and {\it HST}/STIS.\label{carb}}
\vspace*{0.5cm}
\plotone{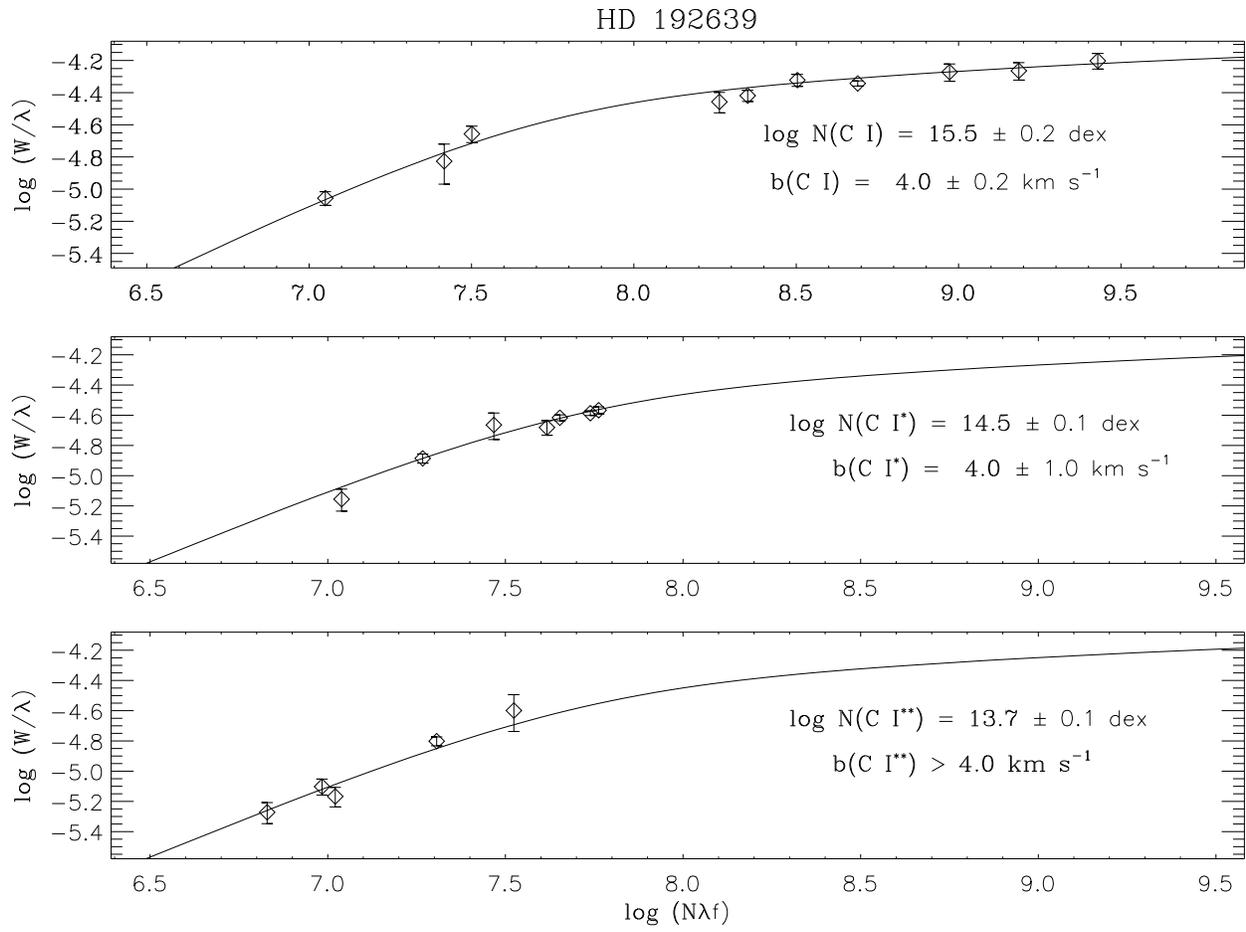}
\end{figure}

\begin{figure}
\caption{Comparison of the gas-phase depletions in the LVCs and the depletions from the ``typical'' cold and warm interstellar gas \citep{welty99} toward HD\,192639. \label{deple}}
\plotone{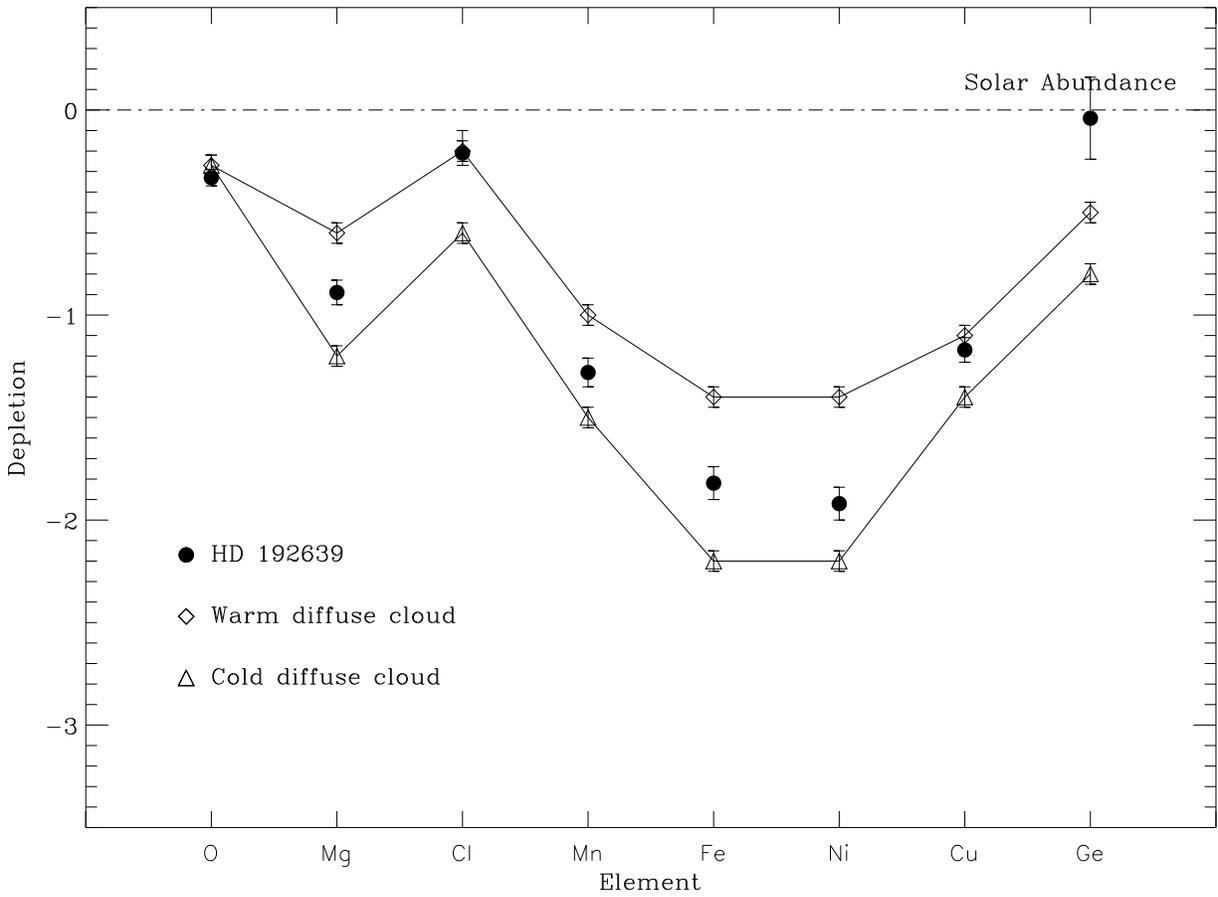}
\end{figure}

\clearpage

\begin{deluxetable}{lcclr}
\tablewidth{0pt}
\tablecaption{\sc Total Equivalent Widths in the LVCs toward HD\,192639.\label{ewlow}}
\tablehead{
\colhead{Species} & \colhead{$\lambda^{a}$} & \colhead{$\log(f\lambda)^{a}$} &\colhead{$W$} & \colhead{Dataset} \\
 & \colhead{(\AA)} & &\colhead{(m\AA)} & }
\startdata
\ion{C}{1} & 1129.195 & 0.998 & 53.8 $\pm$ 4.6 & FUSE\\
           & 1129.317 & $-$0.089$^{b}$ & 16.8 $\pm$ 4.7 & FUSE\\
           & 1155.809 & 0.845 & 44.1 $\pm$ 3.5 & FUSE\\
           & 1192.215 & $-$0.004 & 26.3 $\pm$ 3.1 & STIS\\
           & 1193.995 & 1.184 & 54.0 $\pm$ 2.0 & STIS\\
	   & 1260.735 & 1.681 & 68.6 $\pm$ 8.5 & STIS\\
           & 1270.143 & $-$0.455 & 11.2 $\pm$ 1.1 & STIS\\
           & 1276.482 & 0.758 & 44.6 $\pm$ 6.5 & STIS\\
           & 1280.135 & 1.468 & 68.4 $\pm$ 8.4 & STIS\\
           & 1328.833 & 1.924 & 83.5 $\pm$ 9.3 & STIS\\
\ion{C}{1*} & 1193.678 & 1.085 & 24.9 $\pm$ 2.8 & STIS\\
	    & 1260.926 & 1.231 & 34.2 $\pm$ 1.8 & STIS\\
	    & 1260.996 & 1.122 & 30.5 $\pm$ 1.4 & STIS\\   
            & 1276.749 & 0.507 & 8.9 $\pm$  1.5 & STIS\\
            & 1279.890 & 1.208 & 33.2 $\pm$ 1.2 & STIS\\
            & 1280.404 & 0.736 & 16.7 $\pm$ 1.2 & STIS\\
            & 1280.597 & 0.936 & 27.8 $\pm$ 5.5 & STIS\\
\ion{C}{1**} & 1261.426 & 1.105&  6.7 $\pm$ 1.1 & STIS\\
             & 1261.552 & 1.582 & 19.9 $\pm$ 1.3 & STIS\\
             & 1277.723 & 1.297 & 8.7 $\pm$  1.3 & STIS\\
             & 1280.333 & 1.259 & 10.1 $\pm$ 1.2 & STIS\\ 
             & 1329.577 & 1.800 & 33.5 $\pm$ 9.2 & STIS\\
\ion{O}{1}  & 1355.597 & $-$2.803 & 19.8 $\pm$ 1.1 & STIS \\
\ion{O}{1*} & 1304.857 & 1.830 & 6.6 $\pm$ 0.8 & STIS\\
\ion{O}{1**} & 1306.028& 1.830 & 4.9 $\pm$ 1.0 & STIS\\
\ion{Mg}{1} & 1827.935$^{b}$ & 1.651$^{b}$ & $>$ 89 & IUE\\
\ion{Mg}{2} & 1239.925 & $-$0.116$^{c}$ & 63.0 $\pm$ 1.7 & STIS\\
	    & 1240.395 &$-$0.357$^{c}$ & 48.9 $\pm$ 1.5 & STIS\\
\ion{P}{2}  & 1152.818 & 2.451 &136.7 $\pm$ 16.5 & FUSE   \\
\ion{P}{3} & 1334.813 & 1.576 & 5.4 $\pm$ 1.7 & STIS  \\
\ion{S}{1} & 1295.653 & 2.052$^{d}$ & 27.0 $\pm$ 1.0 & STIS   \\
           & 1296.174 & 1.455$^{d}$ & 9.0 $\pm$ 1.0 & STIS\\
\ion{Cl}{1}& 1004.677 & 2.200 & 62.0$\pm$ 14.0 & FUSE\\
           & 1088.059 & 1.945$^{e}$ & 50.0$\pm$ 15.0  & FUSE\\
           & 1097.370 & 0.985$^{e}$  & 10.0$\pm$ 3.0 & FUSE\\
           & 1347.239 & 2.314$^{e}$ & 70.0$\pm$ 2.0 & STIS\\
\ion{Cl}{2}& 1071.035 & 1.206 & 26.3 $\pm$ 7.4 & FUSE\\
\ion{Ar}{1}& 1048.224 & 2.440 & $>$ 175 & FUSE\\
\ion{Mn}{2} & 1197.184 & 2.415 & 52.5 $\pm$ 3.2 & STIS\\
            & 1201.118 & 2.163 & 39.4 $\pm$ 2.4 & STIS  \\
\ion{Fe}{2} & 1055.261 & 0.898$^{f}$ & 44.3 $\pm$ 8.9 &  FUSE \\
            & 1112.048 & 0.839$^{f}$ & 46.8 $\pm$ 6.6 & FUSE\\
            & 1121.975 & 1.355$^{f}$ & 93.2 $\pm$ 5.9 & FUSE\\
            & 1125.447 & 1.255$^{f}$ & 92.9 $\pm$ 8.3 & FUSE\\
            & 1127.098 & 0.499$^{f}$ & 39.5 $\pm$ 7.2 & FUSE\\
            & 1133.665 & 0.795$^{f}$ & 75.0 $\pm$ 6.0 & FUSE\\
            & 1142.365 & 0.681$^{f}$ & 47.5 $\pm$ 4.6 & FUSE\\
            & 1143.226 & 1.306$^{f}$ & 85.8 $\pm$ 5.4 & FUSE\\      
            & 1144.938 & 2.084$^{f}$ & 137.0 $\pm$ 5.0& FUSE\\
\ion{Ni}{2} & 1317.217 & 2.009$^{g}$ & 49.0 $\pm$ 1.0 & STIS \\ 
\ion{Cu}{2} & 1358.773$^{b}$ & 2.713$^{b}$ & 19.9 $\pm$ 1.0 & STIS \\ 
\ion{Ge}{2} & 1237.059$^{g}$ & 3.183$^{g}$ & 44.9 $\pm$ 3.8 & STIS\\
CO & 1076.034$^{h}$ & 1.864$^{h}$ & 48.0 $\pm$  9.2 & FUSE\\
   & 1087.868$^{i}$ & 2.107$^{i}$ & 55.0 $\pm$ 16.0 & FUSE \\
   & 1344.186$^{i}$ & 0.745$^{i}$ & 5.45 $\pm$ 2.1 & STIS \\
\enddata
\scriptsize{
\tablenotetext{a}{ Vacuum wavelengths and $f$-values from \citet{morton}, unless noted otherwise: (b) \citet{morton91}; (c) \citet{sophia00}; (d) \citet{beideck}; (e) \citet{schectman}; (f) \citet{Howk00b}; (g) \citet{welty99}; (h) \citet{federman} and (i) \citet{morton94}. Uncertainties are 1$\sigma$.}
}
\end{deluxetable}

\begin{deluxetable}{llllll}
\tablewidth{0pt}
\tablecaption{\sc Column densities of the individual components of the LVCs.\label{tab1}}
\tablehead{
\colhead{Species} & \colhead{Comp 1} & \colhead{Comp 2} & \colhead{Comp 3} & \colhead{Comp 4} & \colhead{Comp 5}\\
 \colhead{ $v$(\kms)}& \colhead{$-$18.5}& \colhead{$-$16.5} & \colhead{$-$14.0}& \colhead{$-$10.7} & \colhead{$-$8.4}\\
 \colhead{ $b$(\kms)}& \colhead{0.8} & \colhead{1.1} & \colhead{1.0} & \colhead{1.2} & \colhead{1.1}}
\startdata
\ion{C}{1} & 5.5$\pm$5.0E13 & 6.5$\pm$1.6E14  & 2.7$\pm$0.9E14  & 8.5$\pm$5.9E14  & 1.4$\pm$0.5E14 \\
\ion{C}{1*} & 1.1$\pm$1.0E13  & 1.5$\pm$1.1E14  & 4.3$\pm$2.1E13  & 1.1$\pm$1.0E14  & 2.5$\pm$0.8E13\\
\ion{C}{1**}& 1.3$\pm$1.0E12  & 1.9$\pm$1.2E13  & 6.1$\pm$2.9E12  & 1.3$\pm$0.6E13 & 4.9$\pm$2.3E12  \\
\ion{O}{1} & 2.4$\pm$1.5E17  & 2.7$\pm$1.5E17  &  2.9$\pm$0.2E17  & 3.5$\pm$0.3E17  & 2.3$\pm$1.6E17 \\
\ion{O}{1*}& 5.3$\pm$2.3E11  & 1.4$\pm$0.2E12  & 2.8$\pm$0.4E12  & 2.2$\pm$0.1E12  & $<$5.2E10 \\
\ion{O}{1**}& ----- & 2.1$\pm$0.6E12  & 1.3$\pm$1.0E12  & 1.2$\pm$0.2E12  & -----\\
\ion{Na}{1}$^{a}$ & [1.01E13] & [2.99E13] & [6.10E12]& [1.99E13] & [1.44E13]\\
\ion{Mg}{2}& 2.2$\pm$1.7E15  & 5.5$\pm$0.2E15  & 4.4$\pm$3.5E15  & 3.9$\pm$3.6E15  & 3.4$\pm$1.1E15 \\
\ion{S}{1}& 5.1$\pm$1.3E12  & 8.5$\pm$1.2E12  & 4.7$\pm$2.2E12  & 7.1$\pm$1.0E12 & 3.2$\pm$1.5E12  \\
\ion{K}{1} & 1.0$\pm$0.1E11& 2.5$\pm$0.1E11 & 0.6$\pm$0.1E11& 2.0$\pm$0.2E11& 1.4$\pm$0.2E11\\ 
\ion{Mn}{2} & 8.9$\pm$0.8E12  & 1.5$\pm$0.5E13  & 1.6$\pm$0.2E13 & 2.7$\pm$0.3E13  & 4.7$\pm$0.2E12 \\
\ion{Ni}{2}& 1.3$\pm$1.1E13  & 2.1$\pm$0.5E13  & 1.4$\pm$1.2E13  & 1.3$\pm$1.2E13 &  9.3$\pm$0.4E12\\
\ion{Cu}{2}& 2.7$\pm$1.4E11  & 1.1$\pm$0.3E12  & 1.3$\pm$0.3E12  &1.7$\pm$0.9E12  & 6.7$\pm$3.5E11\\
\ion{Ge}{2}& 8.7$\pm$2.1E11  & 2.1$\pm$1.5E12  & 2.7$\pm$1.2E11 & 1.7$\pm$0.7E13 & -----\\
\enddata
\scriptsize{
\tablenotetext{a}{ Values in brackets were fixed in the fitting. Uncertainties are 1$\sigma$. Column densities are given in cm$^{-2}$. $v$ and $b$ are the velocity and $b$-value for each component, respectively.}
}
\end{deluxetable}

\begin{deluxetable}{llll}
\tablewidth{0pt}
\tablecaption{\sc Column densities in the LVCs toward HD\,192639.\label{low}}
\tablehead{
\colhead{Species} & \colhead{$\log$ N} & \colhead{Methods$^{a}$} & \colhead{Dataset}\\
  & \colhead{(dex)} & & }
\startdata
\ion{C}{1}  & 15.29 $\pm$ 0.08 & COG;AOD;FIT & FUSE+STIS\\
\ion{C}{1*}  &  14.53 $\pm$ 0.08 & COG;AOD;FIT & FUSE+STIS   \\
\ion{C}{1**}  & 13.67 $\pm$ 0.08 & COG;AOD;FIT & FUSE+STIS   \\
\ion{O}{1}   & 18.16 $\pm$ 0.11 &AOD;COG;FIT &FUSE+STIS \\
\ion{O}{1*}   & 12.91 $\pm$ 0.10 &EW;FIT & STIS\\
\ion{O}{1**}  & 12.73 $\pm$ 0.15 &EW;FIT& STIS\\
\ion{Na}{1}   &  13.91 $\pm$ 0.04 & FIT & Optical\\
\ion{Mg}{1}   & $>$14.09 &AOD & IUE\\
\ion{Mg}{2}  &16.20 $\pm$ 0.09 & COG;AOD;FIT & STIS   \\
\ion{P}{2}    & $>$15.01 & COG;FIT&FUSE   \\
\ion{P}{3}    &13.10 $\pm$ 0.40 & AOD & STIS  \\
\ion{S}{1}  &13.44 $\pm$ 0.07 &  COG;AOD;FIT & STIS   \\
\ion{Cl}{1}   & 14.2 - 14.5$^{a}$ & AOD;FIT & STIS+FUSE \\
\ion{Cl}{2}   & 14.24 $\pm$ 0.10 &EW;AOD & FUSE \\
\ion{K}{1}    & 11.88 $\pm$ 0.04 & FIT & Optical\\
\ion{Ar}{1}   & $>$14.80 & FIT & FUSE \\
\ion{Mn}{2}   &13.76 $\pm$ 0.07 & COG;AOD;FIT & STIS  \\
\ion{Fe}{2}  &15.20 $\pm$ 0.20 & COG;FIT & FUSE     \\
\ion{Ni}{2}   &13.84 $\pm$ 0.15 & AOD;FIT & STIS \\ 
\ion{Cu}{2}  &12.61 $\pm$ 0.09 & AOD;FIT & STIS \\ 
\ion{Ge}{2} &13.10 $\pm$ 0.20 & FIT& STIS\\
\Htwo(J=0) &20.40 $\pm$ 0.10 & FIT& FUSE  \\
\Htwo(J=1) &20.50 $\pm$ 0.10 & FIT& FUSE  \\
\Htwo(J=2)   &18.53 $\pm$ 0.10 & FIT& FUSE\\
\Htwo(J=3)   &17.10 $\pm$ 0.15 & FIT& FUSE \\
\Htwo(J=4)  &15.40 $\pm$ 0.15& FIT& FUSE \\
\Htwo(J=5)  &15.02 $\pm$ 0.20& FIT& FUSE \\
CO  & 14.20 $\pm$ 0.20 & COG;AOD;FIT & FUSE+STIS\\
H\,I  &  21.32 $\pm$ 0.12$^{b}$ & & IUE\\
\enddata
\scriptsize{
\tablenotetext{a}{ Indicates the methods used to derive the column densities (N): the curve of growth (COG), Apparent Optical Depth (AOD), Equivalent Width (EW) and the profile fitting code ``Owens'' (FIT). When multiple methods were used, each column density was weighted by its error to produce the weighted means reported here. The associated errors are 1$\sigma$. See chlorine discussion in \S\, 4.6.}
\tablenotetext{b}{ From \citet{Diplas94}}
}
\end{deluxetable}



\clearpage

\begin{deluxetable}{lccr}
\tablewidth{0pt}
\tablecaption{\sc Depletions in the LVCs toward HD\,192639.}
\tablehead{
\colhead{Species} & \colhead{$\log(\rm{X/H})_{\odot}^{a}$} & \colhead{$\log(\rm{X/H})$} & \colhead{Depletion$^{b}$ (1$\sigma$)}}
\startdata
\ion{C}{0} & $-$3.45 & --- & $-$0.40 (0.10)\\
\ion{O}{0} & $-$3.26 & $-$3.46 & $-$0.20 (0.04) \\
\ion{Na}{0} & $-$6.29 & --- & $-$0.60 (0.10) \\
\ion{Mg}{0} & $-$4.42 & $-$5.31 & $-$0.89 (0.06)\\
\ion{S}{0} & $-$4.73 & --- & 0.00 (0.10) \\
\ion{Cl}{0} & $-$6.73 & $-$6.94$^{c}$& $-$0.21($_{-0.06}^{+0.11}$)\\
\ion{K}{0} & $-$7.57 & --- & $-$0.70 (0.10)\\
\ion{Mn}{0} & $-$6.47 & $-$7.75 & $-$1.28 (0.07)\\
\ion{Fe}{0} & $-$4.49 & $-$6.31 & $-$1.82 (0.08)\\
\ion{Ni}{0} & $-$5.75 & $-$7.67 & $-$1.92 (0.08)\\
\ion{Cu}{0} & $-$7.73 & $-$8.90 & $-$1.17 (0.06)\\
\ion{Ge}{0} & $-$8.37 & $-$8.41 & $-$0.04 (0.20) \\
\enddata
\scriptsize{
\tablenotetext{a}{ Solar meteoritic and photospheric (C) abundances from \citet{anders89} and \citet{greve93}; O abundance from \citet{hol01}; logarithmic with H=12.00.}
\tablenotetext{b}{ 1$\sigma$ errors are mentioned in parenthesis. The (---) marks the species for which the depletion was assumed at the value in column (4). For those species a typical error of 0.10 was also assumed.}
\tablenotetext{c}{ Estimate of the abundance and depletion of chlorine using $\log N$({\rm Cl\,I})= 14.3 dex (see \S\, 4.6).\label{deple1}}
}
\end{deluxetable}

\begin{deluxetable}{lcccccc}
\tablewidth{0pt}
\tablecaption{\sc Electron density in the LVCs toward HD\,192639.\label{dens}}
\tablehead{
\colhead{Species} &  \colhead{$\log$ \rm{N}(\ion{X}{1})} & \colhead{$\log$ \rm{N}(\ion{X}{2})$^b$} &\colhead{$\Gamma/\alpha$$^a$ (WJ1)}& \colhead{$n_{e}$ (WJ1)} & \colhead{$\Gamma/\alpha$$^a$ (D)} & \colhead{$n_{e}$ (D)} \\
 &\colhead{(dex)} &\colhead{(dex)} & \colhead{(cm$^{-3}$)} & \colhead{(cm$^{-3}$)} & \colhead{(cm$^{-3}$)} & \colhead{(cm$^{-3}$)}} 
\startdata
\ion{Na}{0} & 13.90 $\pm$ 0.05 & (15.20 $\pm$ 0.10) & 2.2 & 0.112 & 2.6 & 0.132\\
\ion{Mg}{0} & $\ge$14.09 & 16.20 $\pm$ 0.07 & 13.5 & 0.106 & 12.1 & 0.094 \\
\ion{K}{0} & 11.85 $\pm$ 0.02 & (13.95 $\pm$ 0.05) & 10.1 & 0.087 & 11.4 & 0.099 \\
\ion{C}{0} & 15.29 $\pm$ 0.07 & (17.66 $\pm$ 0.05) & 24.0 &0.102 & 40.0 & 0.171 \\
\ion{S}{0} & 13.44 $\pm$ 0.07 & (16.78 $\pm$ 0.03) & 74.0 & 0.034 & 114.0 & 0.052 \\
\enddata
\scriptsize{
\tablenotetext{a}{ $\Gamma/\alpha$ from \citet{aldro73}, \citet{aldro74}, 
\citet{pequi86} and \citet{shull82}. $\alpha$ is calculated for T=100 K. (WJ1) and (D) refers to the average radiation field as in \citet{witt} and \citet{draine}, respectively.}
\tablenotetext{b}{ The parenthesis mark the species for which the depletion was assumed (see Table \ref{deple1}). Errors are 1$\sigma$.}
}
\end{deluxetable}



\clearpage

\begin{deluxetable}{lclr}
\tablewidth{250pt}
\tablecaption{\sc Column densities of the HVCs toward HD\,192639.\label{high}}
\tablehead{
\colhead{Species} & \colhead{$\lambda$} & \colhead{$\log$ N} & \colhead{Dataset} \\
 & \colhead{(\AA)} & \colhead{(dex)$^{*}$} & \\
}
\startdata
\ion{C}{3} & 977.020 & $>$ 13.50 & FUSE\\
\ion{C}{4} & 1548.188 & 13.79$^{+0.08}_{-0.15}$ &IUE$^{a}$\\
\ion{N}{1} & 1200.230 & 12.77$^{+0.29}_{-0.48}$ & STIS \\
\ion{N}{2} & 1084.584 & [15.79] & FUSE+STIS   \\
\ion{N}{2*} & 1084.584 &14.40$^{+0.14}_{-0.20}$ & FUSE   \\
\ion{Al}{3} & 1862.790 & 13.02$^{+0.14}_{-0.15}$&IUE$^{a}$\\
\ion{Si}{2} & 1020.667 &14.62$^{+0.12}_{-0.17}$ & FUSE\\
\ion{Si}{2*} & 1264.738 &12.31$^{+0.03}_{-0.04}$ & STIS    \\
\ion{Si}{4} & 1402.770 &13.39$^{+0.14}_{-0.15}$ & IUE$^{a}$\\
\ion{S}{2} & 1250.535 &14.54$^{+0.04}_{-0.04}$ & STIS   \\
\ion{S}{2} & 1253.811 &14.28$^{+0.04}_{-0.05}$ & STIS   \\
\ion{S}{2} & 1259.519 &14.50$^{+0.02}_{-0.06}$ & STIS   \\
\ion{Fe}{2}& 2344.213 & 12.95$^{+0.17}_{-0.26}$ &IUE$^{a}$\\
\ion{Fe}{2}& 2374.458 & 12.95$^{+0.17}_{-0.26}$ & IUE$^{a}$\\
\enddata
\scriptsize{
\tablenotetext{*}{ All {\it FUSE} and STIS column densities were derived using the AOD method in the velocity range [$-$100, $-$50] \kms. Errors are 1$\sigma$. Values in brackets are estimates (see \S\, 5.3).}
\tablenotetext{a}{ From \citet{phi84}}}
\end{deluxetable}




\end{document}